\documentclass[twocolumn,showpacs,preprintnumbers,amsmath,amssymb,aps,prl]{revtex4-1}

\usepackage{graphicx}
\usepackage{dcolumn}
\usepackage[columnwise]{lineno}
\usepackage{bm}
\usepackage{epstopdf}
\usepackage{psfrag}
\usepackage[T1]{fontenc}
\usepackage{hyperref}
\usepackage{color}
\usepackage{tabularx}
\newcommand\numberthis{\addtocounter{equation}{1}\tag{\theequation}}

\newcolumntype{Y}{>{\centering\arraybackslash}X}
\usepackage{titlesec}
\titleformat{\section}[hang]{\normalfont\large\bfseries}{\thesection.}{\noindent}{}
\titlespacing\section{0pt}{12pt plus 4pt minus 2pt}{-\parskip}\relax
\begin{document}
%\linenumbers
%\linenumbersep=5pt

%\runningpagewiselinenumbers
\preprint{}

\title{The origin of the lattice thermal conductivity enhancement at the ferroelectric phase transition in GeTe}
\author{{\DJ}or{\dj}e Dangi{\'c}\textsuperscript{1,2}}
\email{djordje.dangic@tyndall.ie}
\author{Olle Hellman\textsuperscript{3}}
\author{Stephen Fahy\textsuperscript{1,2}}
\author{Ivana Savi\'c\textsuperscript{2}}
\email{ivana.savic@tyndall.ie}
\affiliation{\textsuperscript{\normalfont{1}}Department of Physics, University College Cork, College Road, Cork T12 K8AF, Ireland}
\affiliation{\textsuperscript{\normalfont{2}}Tyndall National Institute, Dyke Parade, Cork T12 R5CP Ireland}
\affiliation{\textsuperscript{\normalfont{3}}Department of Physics, Chemistry and Biology (IFM), Link{\"o}ping University, SE-581 83, Link{\"o}ping, Sweden}

\date{\today}

\begin{abstract}

The proximity to structural phase transitions in IV-VI thermoelectric materials is one of the main reasons for their large phonon anharmonicity and intrinsically low lattice thermal conductivity $\kappa$. However, the $\kappa$ of GeTe increases at the ferroelectric phase transition near $700$ K. Using first-principles calculations with the temperature dependent effective potential method, we show that this rise in $\kappa$ is the consequence of negative thermal expansion in the rhombohedral phase and increase in the phonon lifetimes in the high-symmetry phase. Negative thermal expansion increases phonon group velocities, which counteracts enhanced anharmonicity of phonon modes and boosts $\kappa$ close to the phase transition in the rhombohedral phase. A drastic decrease in the anharmonic force constants in the cubic phase increases the phonon lifetimes and $\kappa$. Strong anharmonicity near the phase transition induces non-Lorentzian shapes of the phonon power spectra. To account for these effects, we implement a novel method of calculating $\kappa$ based on the Green-Kubo approach and find that the Boltzmann transport equation underestimates $\kappa$ near the phase transition. Our findings elucidate the influence of structural phase transitions on $\kappa$ and provide guidance for design of better thermoelectric materials.

\end{abstract}

\pacs{65.40.De, 63.20.-e, 64.60.-i}

\maketitle

\section{Introduction.}
Reducing the lattice thermal conductivity $\kappa$ is one of the most successful ways of improving the efficiency of thermoelectric materials \cite{lowkapa1, lowkapa2, lowkapa3, lowkapa4, lowkapa5, lowkapa6, lowkapa7}. Many of the best thermoelectric materials have intrinsically low lattice thermal conductivity. This is usually related to the large anharmonicity of phonon modes, which can stem from weak bonding, as in van der Waals materials \cite{lowkapa2,lowkapa5, SnSkapa, BiTekapa1, BiSekapa, BiTekapa2}, or rattling modes \cite{Rattle0, Rattle1, Rattle1.5, Rattle2, Rattle3, Rattle4}. Another source of large anharmonicity in good thermoelectric materials can be the proximity to structural phase transitions, as in the case of IV-VI materials \cite{DelairePbTe, Shiga2012, PbTeSpect, PbTeOlle, Aoife1, Aoife2, GeTeChan}. For example, germanium telluride (GeTe) undergoes the ferroelectric phase transition at $\sim 700$ K, and has intrinsically low lattice thermal conductivity and high thermoelectric efficiency \cite{GeTe1, GeTe2, GeTe3}.

Recent computational work has predicted that driving IV-VI materials closer to the ferroelectric phase transition via strain or alloying can lead to a drastically lower lattice thermal conductivity \cite{Aoife1, Aoife2}. Under the assumption of the displacive phase transition, it was found that coupling between soft transverse optical (TO) modes and the heat carrying acoustic modes is the main reason for the $\kappa$ reduction. At the displacive transition, the frequency of the soft TO mode collapses, becoming effectively zero. Since scattering rates are inversely proportional to phonon frequencies, the lifetimes of the acoustic modes that couple to soft TO modes decrease dramatically, leading to a considerable $\kappa$ reduction \cite{Aoife1, Aoife2}.

Surprisingly, experimental studies have shown that the lattice thermal conductivity increases at the ferroelectric phase transition in GeTe \cite{GeTe1, GeTe2, GeTe3, GeTe4, GeTe5, GeTe6, GeTe7, GeTe8}. This is at odds with measurements in some other materials going through ferroelectric phase transitions, where a significant decrease in $\kappa$ is observed \cite{ThermCondPT}. The reason for the anomalous behaviour of $\kappa$ at the phase transition in GeTe remains unknown. Understanding the microscopic origin of the $\kappa$ increase at the ferroelectric phase transition in GeTe may lead to design of improved thermoelectric materials.

Here we study how driving GeTe near the ferroelectric phase transition via temperature affects its lattice thermal conductivity, making no assumptions about the nature of the phase transition. Unlike the previous work \cite{Aoife1, Aoife2}, we calculate interatomic force constants at different temperatures using the state-of-the art, temperature dependent effective potentials (TDEP) method \cite{TDEP1, TDEP2, TDEP3}. We find that the increase of the lattice thermal conductivity of GeTe at the phase transition in the rhombohedral phase comes from negative thermal expansion that enhances the phonon group velocities. In the cubic phase the phonon lifetimes increase, leading to even more substantial increase of $\kappa$. Large anharmonicity of phonon modes minimizes the phonon lifetimes at the phase transition in the rhombohedral phase and leads to non-Lorentzian power spectra of phonon modes. We implement a new method of calculating $\kappa$ that includes these non-Lorentzian lineshapes of the phonon power spectra near the phase transition. This approach further increases $\kappa$ at the phase transition, which can be attributed to further softening of the phonon frequencies due to phonon-phonon interaction. \\

\section{Results and discussion.}
\textbf{TO mode softening at the phase transition.}
GeTe is a ferroelectric material, exhibiting a spontaneous polarization below $600 - 700$~K \cite{FerGeTe1, FerGeTe2, FerGeTe3, FerGeTe4} (the critical temperature strongly depends on the free charge carrier concentration). This occurs due to a slight offset of the Te sublattice along one of the body diagonals of the rocksalt structure. At temperatures higher than $600 - 700$ K, GeTe transforms to the rocksalt structure, losing its ferroelectric nature \cite{main, newmain, GeTeOrderDisorder, GeTeOrderDisorder2}. Below $600 - 700$~K, the GeTe structure can be described by the following set of lattice vectors:
\begin{align*}
\vec{r}_{1} &= a(b,0,c), \\
\vec{r}_{2} &= a(-\frac{b}{2},\frac{b\sqrt{3}}{2},c), \numberthis \label{eq1} \\
\vec{r}_{3} &= a(-\frac{b}{2},-\frac{b\sqrt{3}}{2},c),
\end{align*}
where $a$ is the lattice constant, $b = \sqrt{2(1-\cos\theta)/3}$, $c = \sqrt{(1+2\cos\theta)/3}$, and $\theta$ is the angle between the primitive lattice vectors. The atomic positions in this structure are taken to be: Ge (0.0,0.0,0.0) and Te (0.5 + $\mu$, 0.5 + $\mu$, 0.5 + $\mu$) in reduced coordinates. If the phase transition is displacive, as assumed in Refs.~\cite{Aoife1, Aoife2}, the angle $\theta$ becomes 60$^{\circ}$ at the phase transition, while the interatomic displacement parameter $\mu$ becomes zero \cite{main, newmain}. In this type of the phase transition, the TDEP frequency of the soft mode also collapses to zero \cite{SoftmodeGeTe, landau}. Here we define the TDEP frequency of the phonon mode as the square root of the eigenvalue of the dynamical matrix for that phonon mode at a certain temperature. In the order-disorder phase transition, both the TDEP frequency of the soft mode and the local interatomic displacement are non-zero \cite{GeTeOrderDisorder, GeTeOrderDisorder2, DispvsOrder}. It is still under debate which type of the phase transition occurs in GeTe \cite{SoftmodeGeTe, main, newmain, GeTeOrderDisorder, GeTeOrderDisorder2}.

%Our calculations using the TDEP method of computing temperature dependent interatomic force constants support the order-disorder picture \cite{GeTeOrderDisorder, GeTeOrderDisorder2}. Fig.~\ref{fig1} shows the harmonic frequencies of the two TO modes at the zone centre. The phase transition temperature obtained in our calculations is $T_{C} = 707$~K (see Methods). The harmonic frequencies of both TO modes are strongly renormalised by thermal disorder. Most importantly, their values are non-zero very near the phase transition, indicating that the phase transition in GeTe is of the order-disorder type.

\begin{figure}[h!]
\begin{center}
\includegraphics[width=0.4\textwidth]{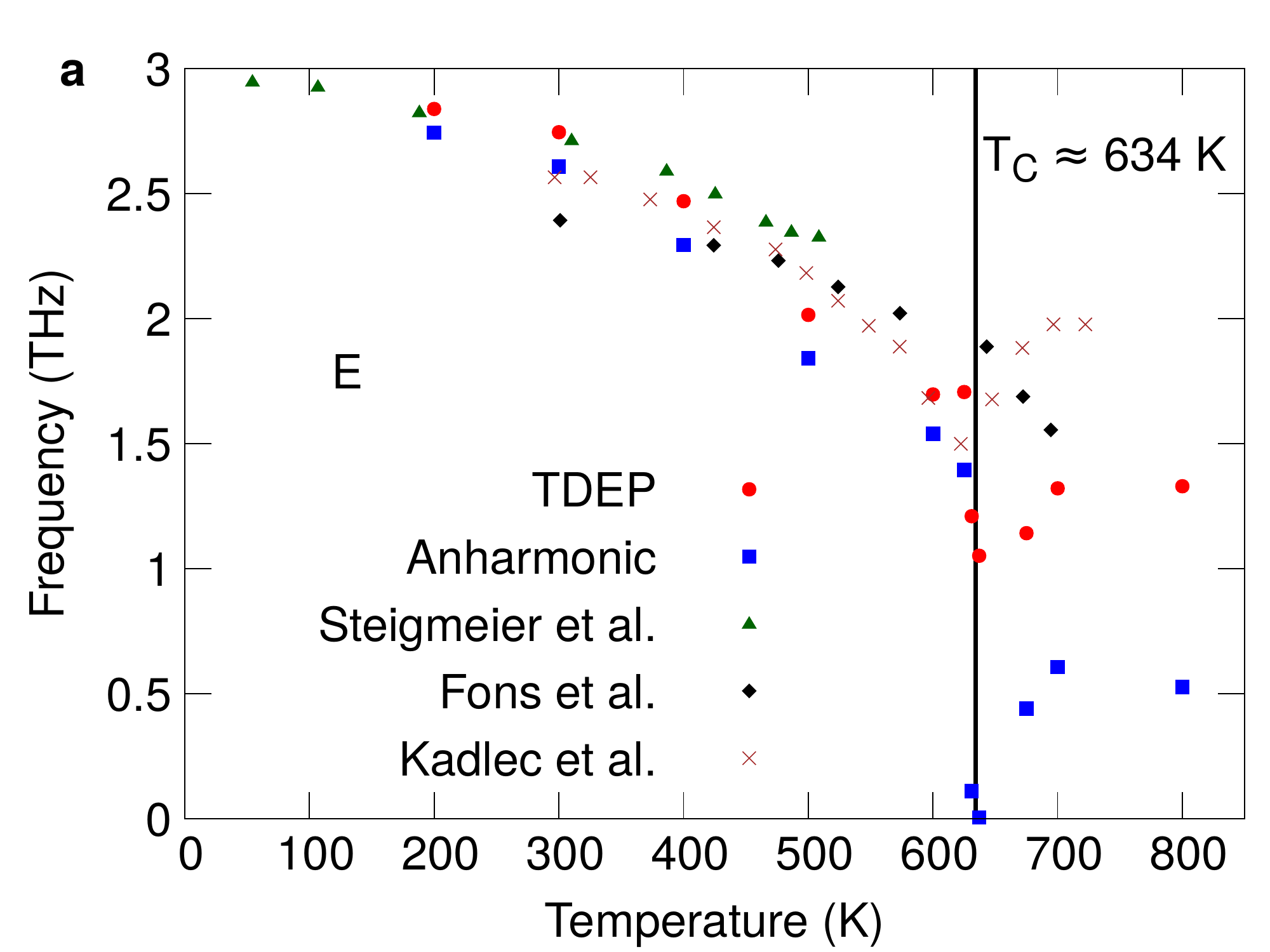}
\includegraphics[width=0.4\textwidth]{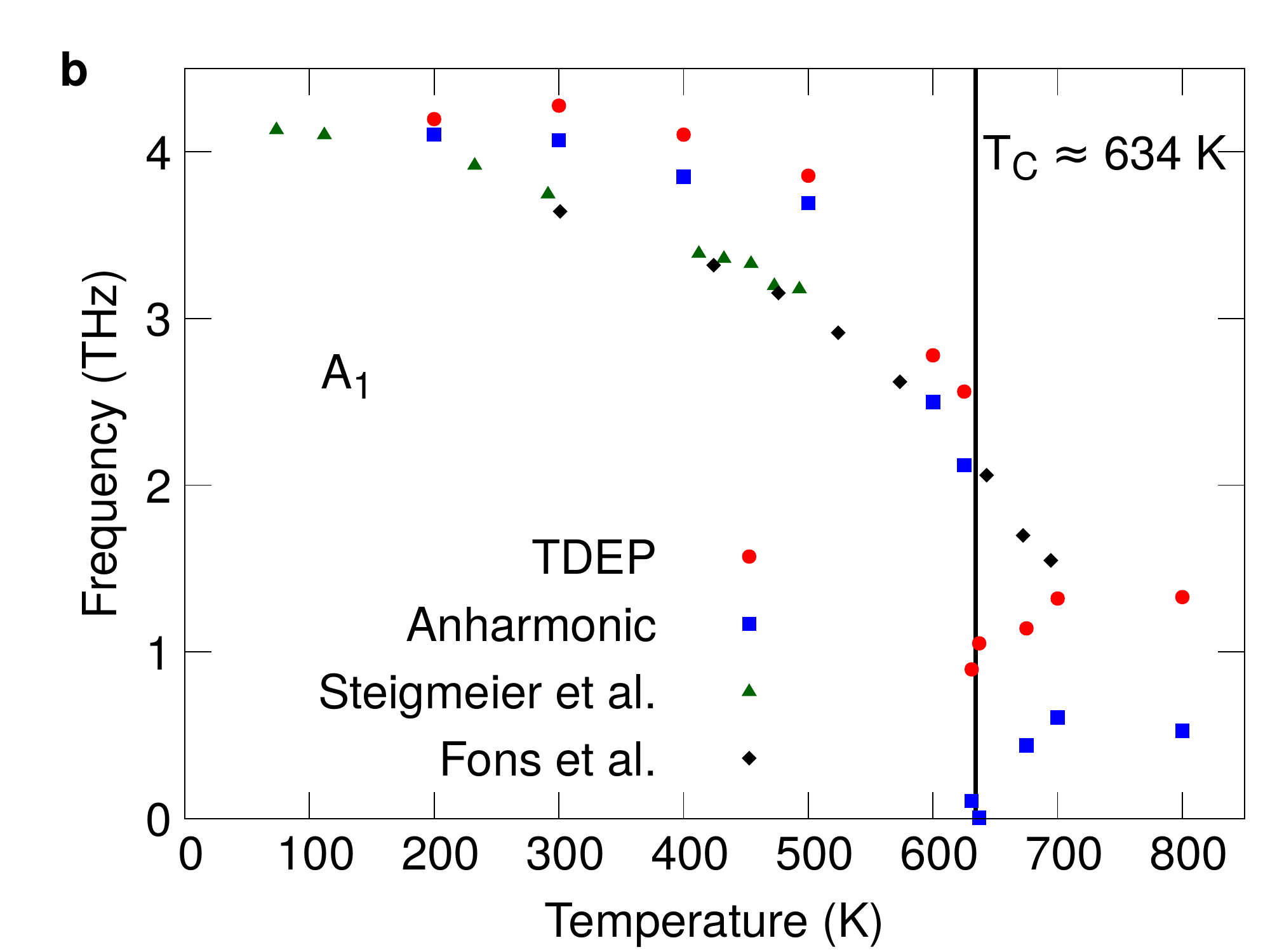}
\caption{\textbf{Temperature dependence of the zone centre optical phonon frequencies.} \textbf{a} The lower transverse optical (TO) mode (E) and \textbf{b} the higher TO mode (A$_1$) at the zone centre as we approach $\vec{q} = 0$ along the $X-\Gamma$ direction, perpendicular to the trigonal axis. Red circles and blue squares represent the calculated temperature dependent effective potential (TDEP) and anharmonic phonon frequencies, while other symbols represent the experimental results from Refs.~\cite{SoftmodeGeTe, Kadlec, GeTeOrderDisorder}. The TDEP frequency is the square root of the eigenvalue of the dynamical matrix at a particular temperature, while the anharmonic frequency is the peak of the phonon mode power spectrum. The black vertical line represents the phase boundary between the rhombohedral and rocksalt structures in our calculations ($\approx 634$ K).}
\label{fig1}
\end{center}
\end{figure}

We have calculated the phonon dispersion of GeTe at different temperatures using the TDEP method and neglecting LO/TO splitting due to high hole concentration in experimental GeTe samples. The TDEP method allows us to calculate TDEP frequencies of phonon modes at different temperatures. In these calculations, we used the structural parameters calculated at appropriate temperatures (see the Computational Methods Section). Fig.~\ref{fig1} shows the TDEP frequencies of the two TO modes at the Brillouin zone centre in GeTe as we approach $\Gamma$ from the $\Gamma-X$ direction (corresponding to the non-degenerate A$_1$ mode and the E mode which is degenerate with the longitudinal optical (LO) mode). The phase transition temperature obtained in our calculations is $T_{C} \approx 634$~K (see Computational Methods). The TDEP frequencies of both phonon modes are strongly renormalized by anharmonic interaction and lattice thermal expansion. The TDEP frequencies of these phonon modes soften drastically at the phase transition, but they do not become zero, indicating that the phase transition in GeTe might not be of the displacive type. 

The observed phonon frequency in experiments is not the TDEP frequency, but what we will call in the rest of the paper the anharmonic frequency i.e. the peak of the phonon mode power spectrum. Blue squares in Fig. \ref{fig1} represent the computed anharmonic frequencies of optical modes at the zone centre, which do fall to zero at the phase transition, in contrast to the TDEP frequencies. Our results thus suggest that the observation of phonon mode softening is not a conclusive proof of the displacive type of the phase transition, as previously argued in the case of GeTe \cite{SoftmodeGeTe} (see Supplementary Note 1 for more details). The calculated phonon frequencies (both TDEP and anharmonic) are very similar in the two different phases for the temperatures closest to the phase transition (at 631 K and 637 K).

Our computed anharmonic optical frequencies at the zone centre agree fairly well with those measured in experiments, see Fig.~\ref{fig1}. This agreement highlights the accuracy of the TDEP method even for the challenging cases of materials undergoing structural phase transitions. We note that the critical temperature in Ref.~\cite{SoftmodeGeTe} is around 600 K, in Ref.~\cite{GeTeOrderDisorder} approximately 700 K and in Ref.~\cite{Kadlec} 650 K. This difference in the calculated and measured critical temperature is expected since the critical temperature strongly depends on the number of free charge carriers \cite{Abrikosov, Marchenkov1994}. 

\textbf{Phonon spectral function.} Harmonic (and/or TDEP) frequencies are a valid description of lattice dynamics only in the absence of phonon-phonon interaction. In inelastic neutron scattering experiments that measure phonon spectral functions, harmonic phonons would produce zero linewidth signals, revealing infinitely long lived quasiparticles. However, in real materials phonons interact with each other, thus broadening phonon spectral functions, with linewidths inversely proportional to phonon lifetimes. This effect can be described using the concept of phonon self-energy that quantifies the strength of phonon-phonon interaction \cite{Maradudin, Cowley, Kokadee}.  

The probability of an incoming neutron to interact with a phonon system acquiring/losing energy $\hbar\Omega$ and momentum $\hbar\vec{q}$ is proportional to the spectral function \cite{Maradudin, Cowley, Kokadee, Cowley_Raman}:
\begin{align*}
S(\vec{q}, \Omega) & \sim \sum _{s} \left\langle u_{\vec{q}, s} ^{\dagger} u_{\vec{q}, s}\right\rangle = \sum_{s} \frac{4\omega _{\vec{q},s} ^2}{\pi (e^{\beta \hbar \Omega} - 1)} \times \\
& \frac{\Gamma _{\vec{q},s} (\Omega)}{(\Omega ^2 - \omega _{\vec{q},s} ^2 - 2\omega _{\vec{q},s}\Delta _{\vec{q},s}(\Omega))^2 +4\omega _{\vec{q},s} ^2\Gamma _{\vec{q},s} ^2(\Omega)}, \numberthis
\label{eq2}
\end{align*}
where $u_{\vec{q},s}$ is the phonon displacement operator, $\omega _{\vec{q},s}$ is the TDEP frequency of the phonon mode with wave vector $\vec{q}$ and phonon branch $s$, $\beta$ is $1/k_{B}T$ with $k_{B}$ being the Boltzmann constant and $T$ the temperature, and $\hbar$ is the reduced Planck constant. $\Delta _{\vec{q}, s}$ and $\Gamma _{\vec{q}, s}$ are the real and imaginary part of the phonon-self energy, respectively. In a weakly anharmonic material, this expression can be reduced to a Lorentzian with a half-width of $\Gamma _{\vec{q},s}$ and the position of the peak at $\omega _{\vec{q}, s} + \Delta _{\vec{q},s}$. In this case the phonon lifetime can be calculated as $\tau _{\vec{q},s} = \frac{1}{2\Gamma _{\vec{q},s}}$. However, in materials with strong anharmonicity, phonon spectral functions can exhibit exotic behaviour with satellite peaks, shoulders etc. \cite{DelairePbTe, PbTeSpect, DelaireSnSe, IonPbTe, IonSnSe, IonSnS}. 

\begin{figure}
\begin{center}
\includegraphics[width=0.40\textwidth]{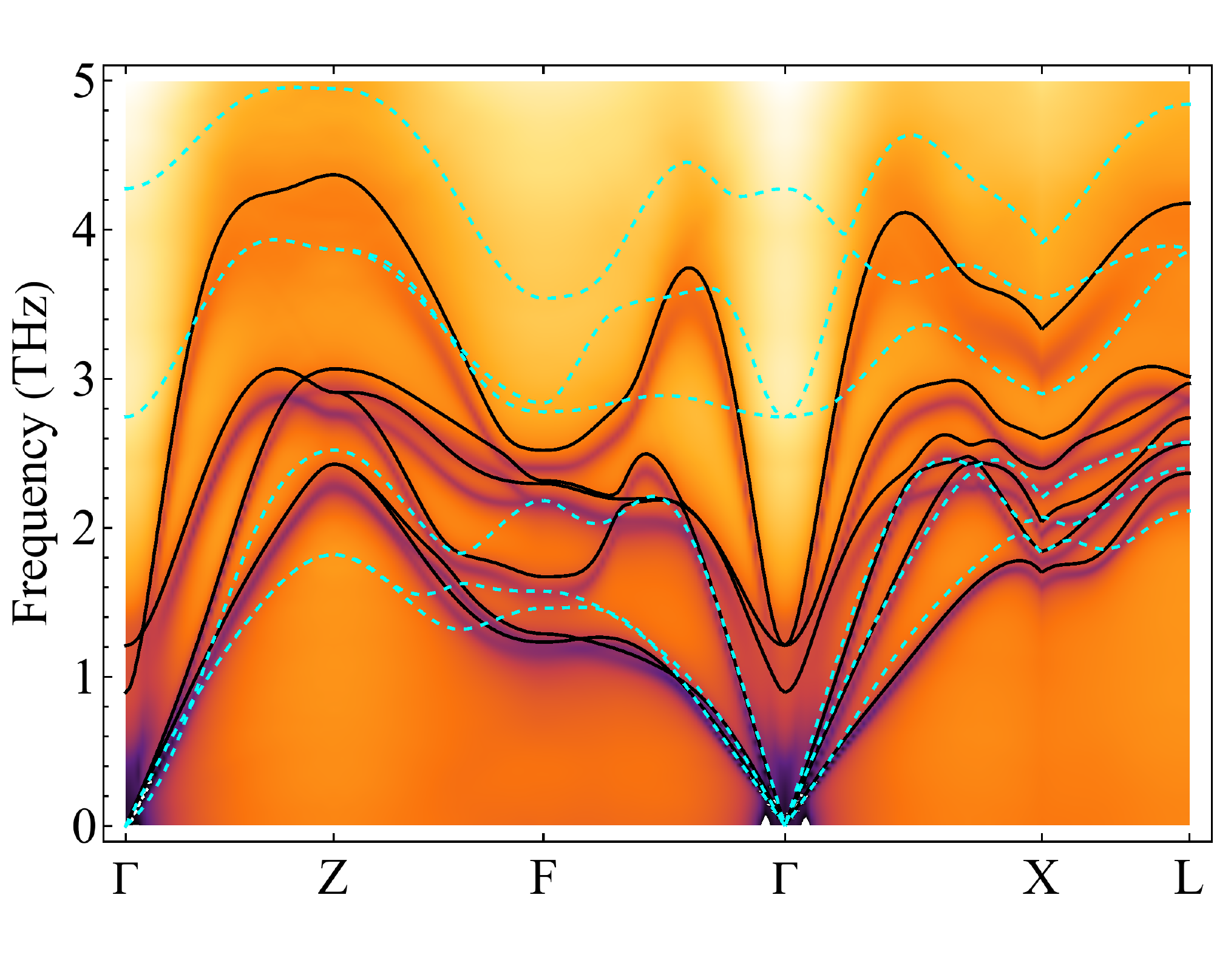}
\caption{\textbf{Phonon spectral function along high symmetry lines in GeTe at the phase transition.} The spectral function is calculated using Eq.~\ref{eq2}. Black lines represent the temperature dependent effective potential (TDEP) frequencies calculated at 631 K (close to the critical temperature 634 K) and dashed cyan lines are the TDEP phonon band structure at 300 K.}
\label{fig2}
\end{center}
\end{figure}

First we show the phonon spectral function of GeTe near the phase transition in the rhombohedral phase (631 K) along a high symmetry path, see Fig.~\ref{fig2}. The black lines represent the TDEP frequencies $\omega _{\vec{q},s}$ calculated at 631 K and the cyan dashed lines are the TDEP frequencies at 300 K. The TDEP frequencies of optical modes soften from 300 to 631 K. The acoustic mode frequencies soften as well in the F - $\Gamma$ and the in plane ($\Gamma$ - X) directions. This is the consequence of the overall softening of the second order force constants with temperature. On the other hand, the acoustic modes along the $\Gamma$ - Z direction (the direction along the trigonal axis) stiffen because of the negative thermal expansion, as we will discuss in the next section. The spectral function shows further softening of the phonon frequencies due to anharmonic phonon-phonon interaction. As expected, the optical phonons at $\Gamma$ can not be clearly resolved in the graph of the phonon spectral function. Supplementary Note 2 shows the phonon band structure in the cubic phase near the phase transition, at 637 K.  

\begin{figure}
\begin{center}
\includegraphics[width=0.45\textwidth]{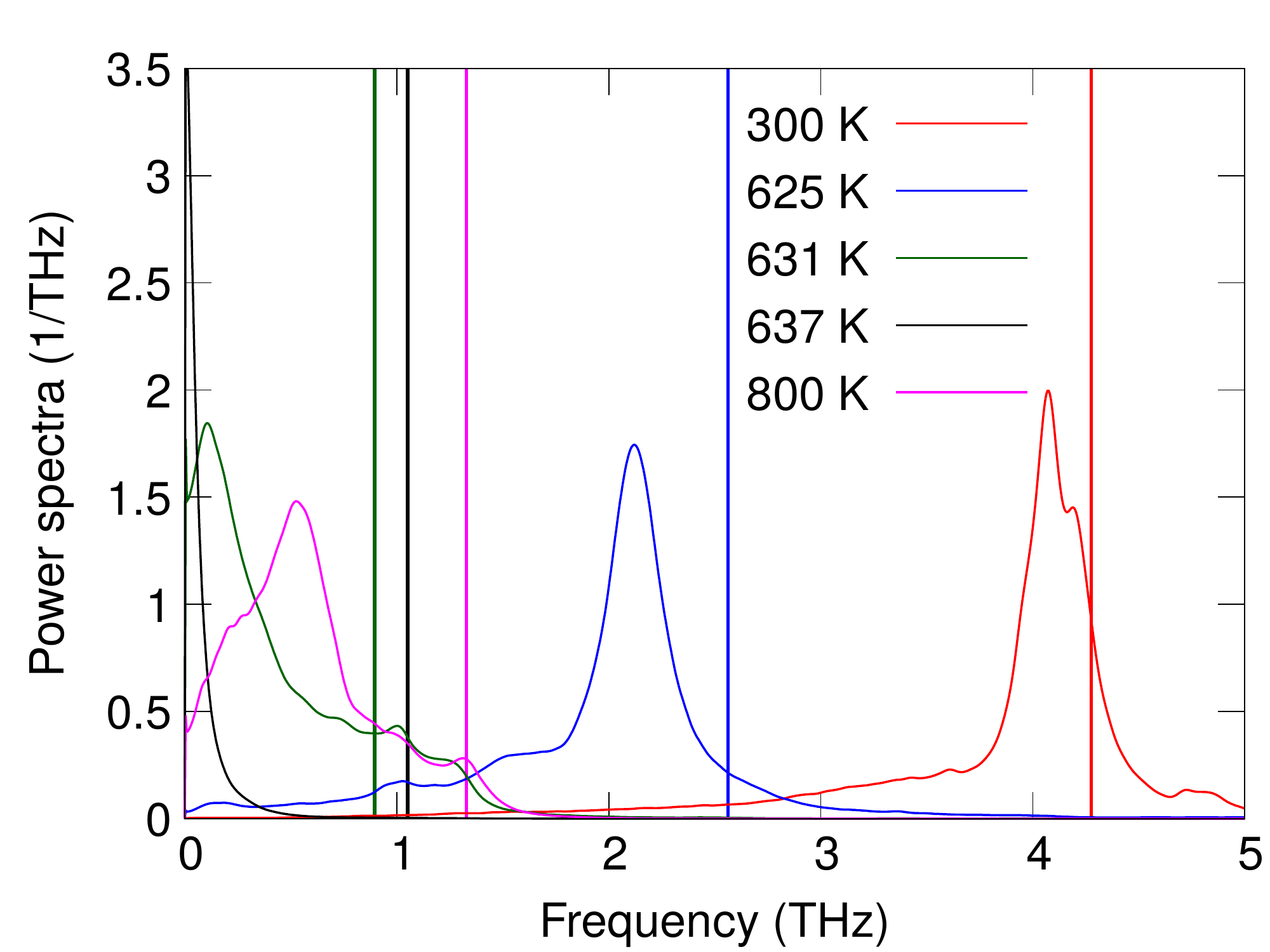}
\caption{\textbf{Spectral function for the zone centre A$_1$ mode in GeTe at different temperatures.} Vertical lines represent the values of the TDEP A$_1$($\Gamma$) frequencies at different temperatures. A drastic deviation from the Lorentzian shape is evident for the entire temperature range and most prominent in the vicinity of the phase transition ($634$~K).}
\label{fig3}
\end{center}
\end{figure}

Figure~\ref{fig3} shows the unusual features of the spectral function for the A$_1$ mode of GeTe at the zone centre for several different temperatures. A non-Lorentzian behavior of the phonon spectral function is evident even at 300 K, very far from the phase transition. The broadening of the power spectrum is large, revealing the short lifetime of this phonon mode. The distortion of the power spectrum is stronger at 625 K, with a very large shift of the peak of the spectral function compared to the TDEP frequency. For temperatures near the phase transition, the power spectrum peaks around 0 THz as expected at the phase transition (see Fig.~\ref{fig1}). At temperatures higher than the phase transition temperature, the peak of the power spectrum is at non-zero frequencies, but is still strongly renormalized compared to the TDEP frequency.

Non-Lorentzian shapes of the phonon power spectra of the optical modes in GeTe at the phase transition are the consequence of the coupling of these phonon modes to the entire phonon bath, rather than coupling to specific phonons. We test this by calculating the power spectrum of the A$_1$ phonon mode disregarding the coupling to specific phonon branches. We find that the change in the power spectrum does not substantially vary depending which phonon branch we disregard. A similar behaviour can be seen in the spectral function of the E mode (see Supplementary Note 3).

\begin{figure}
\begin{center}
\includegraphics[width=0.45\textwidth]{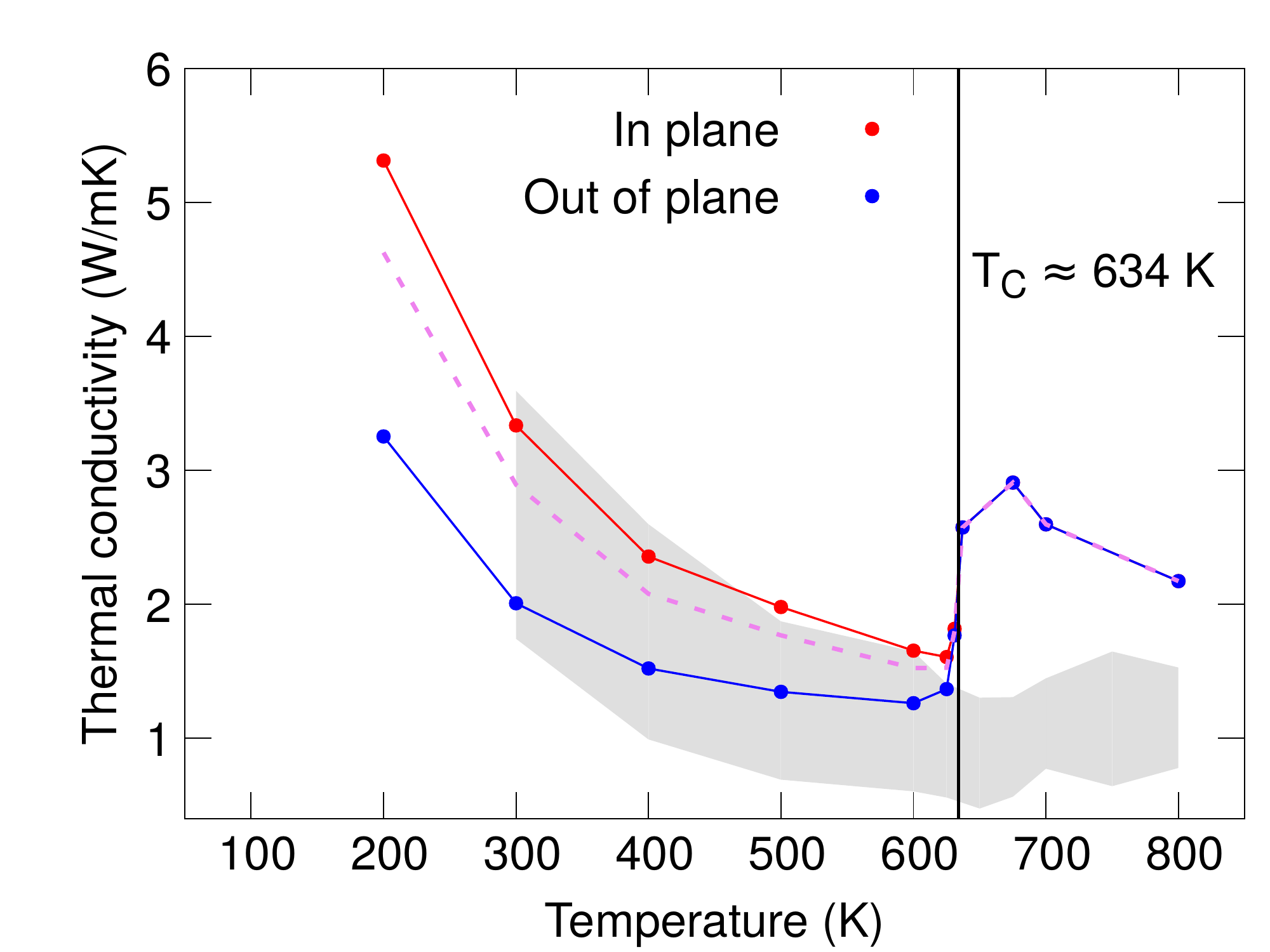}
\caption{\textbf{Temperature dependence of the lattice thermal conductivity of GeTe.} Red and blue lines represent the calculated values using the Boltzmann transport approach in the directions perpendicular and parallel to the trigonal [111] axis, respectively. Dashed line represents the average values of the computed lattice thermal conductivity. Grey region represents the experimental results collected from Refs.~\cite{GeTe1, GeTe2, GeTe3, GeTe4, GeTe5, GeTe6, GeTe7, GeTe8}.}
\label{fig4}
\end{center}
\end{figure}

\textbf{Lattice thermal conductivity in the Boltzmann transport approach.} Next we calculate the lattice thermal conductivity of GeTe for a range of temperatures including both rhombohedral and rocksalt phases (see Fig.~\ref{fig4}), combining the TDEP method with the Boltzmann transport approach. Overall, the lattice thermal conductivity is inversely proportional to temperature, as a result of the linear dependence of phonon populations with temperature above the Debye temperature (~200 K for GeTe). The calculated $\kappa$ deviates from the 1/T law near the phase transition, where there is a large $\kappa$ increase at the phase transition and in the cubic phase. At high temperatures, the $\kappa$ in the cubic phase regains the 1/T dependence.

There is an anisotropy in the lattice thermal conductivity in the rhombohedral phase (Fig.~\ref{fig4}), as a consequence of the van der Waals gaps formed due to the Te sublattice offset. The direction perpendicular to the van der Waals gaps (i.e. parallel to the trigonal [111] axis) has weaker bonding, leading to the lower phonon group velocities and $\kappa$ in that direction. Anisotropy of the lattice thermal conductivity disappears close to the phase transition and is not present in the rocksalt phase. 

The agreement between the computed and experimental $\kappa$ values is very good in the whole temperature range of the rhombohedral phase, especially if we consider the average $\kappa$ (dashed line in Fig.~\ref{fig4}). Almost all experiments show an increase in the lattice thermal conductivity in the vicinity of the phase transition \cite{GeTe1, GeTe2, GeTe3, GeTe4, GeTe5, GeTe6, GeTe7, GeTe8}, similarly to our results. The discrepancy between our results and experiments increases as we get closer to the phase transition and particularly in the cubic phase. Our results consistently overestimate $\kappa$ compared to experiments. In the rhombohedral phase we would expect scattering from lattice imperfections, such as ferroelectric domain walls \cite{OurDW, GeTeDW1, GeTeDW2, GeTeDW3, GeTeDW4}, to further reduce the computed lattice thermal conductivity, but this is not the case in the cubic phase. 

The differences between our calculated $\kappa$ values and experiments could also arise from the fact that GeTe has a large number of Ge vacancies. Several recent publications that report calculations of $\kappa$ in GeTe stress the importance of including point defect scattering in the calculation~\cite{GeTeChan, Campi}. Additionally, the experimental investigation of the phase transition in GeTe~\cite{GeTeBo} noted a huge increase of Ge vacancies in the cubic phase. These point defects scatter higher frequency phonons more effectively ~\cite{Callaway}. Considering this and the fact that the main contribution to the lattice thermal conductivity shifts to the phonons with higher frequencies at higher temperatures (see Supplementary Note 4), we conclude that the lack of point defect scattering in our calculations might be one of the possible reasons for the larger discrepancy in $\kappa$ in the cubic phase between our calculated values and experiments. 

The discrepancies between theory and experiment could also stem from omitting the higher order terms in the Taylor expansion of the interatomic forces (we include the second and third order terms only). However, to the lowest approximation, the fourth order anharmonic terms would only affect the real part of the self-energy \cite{Cowley} and not the imaginary part, which would mean that the phonon lifetimes should remain unchanged. Additionally, since the force constants in TDEP are obtained through a fitting procedure, the fourth order force constants should be smaller than the third order ones. We checked the influence of the fourth order anharmonicity on the real part of the self-energy and found that it is small even for highly anharmonic soft modes at the phase transition. Therefore, higher order anharmonicity is not the reason for the differences in the calculated and experimental $\kappa$ values.

Experimental values of lattice thermal conductivity are usually extracted from the total thermal conductivity measurements using the Wiedemann-Franz law to eliminate the electronic contribution to the thermal conductivity, whose validity at structural phase transitions is not well understood. Additionally, most references use the single parabolic band Kane model to extract the Lorenz factor from measurements of the Seebeck coefficient, which is not appropriate in GeTe due to the intrinsically complicated Fermi surface \cite{GeTe7}. Such lattice thermal conductivity values can differ widely near structural phase transitions, and sometimes an increase in the total thermal conductivity is assigned to the electronic contribution. Here we show that the increase in the thermal conductivity of GeTe at the phase transition, at least partially, comes from the lattice thermal conductivity. The difference between our theoretical and experimental results in the cubic phase might partially be due to an inaccurate estimation of the electronic contribution to the total thermal conductivity in experiments. Measuring the thermal conductivity of GeTe in an applied magnetic field (to exclude the electronic thermal conductivity) would test our predictions of the increased lattice thermal conductivity at the phase transition.

To understand the anomalous behaviour of $\kappa$ near the phase transition, we calculate the average phonon lifetimes and group velocities at different temperatures, Fig.~\ref{fig3}. For example, the average values of the phonon lifetimes in the vicinity of the phonon frequency $\omega _{0}$ are given as:
\begin{align*}
\bar{\tau}(\omega _{0}) = \frac{\sum _{\lambda} \tau (\omega _{\lambda})\exp(-\frac{(\omega _{0} - \omega _{\lambda})^2}{\sigma ^2})}{\sum _{\lambda} \exp(-\frac{(\omega _{0} - \omega _{\lambda})^2}{\sigma ^2})}, \numberthis
\label{eq_av_lifetimes}
\end{align*}
where the sum goes over all phonon modes $\lambda$, and $\sigma$ is the smearing parameter taken to be $\sigma = \omega _{\text{D}}/(N+1)$, where $\omega _{\text{D}}$ is the Debye frequency and $N$ is the number of $\omega _{0}$ frequencies.

\begin{figure}
\begin{center}
\includegraphics[width=0.45\textwidth]{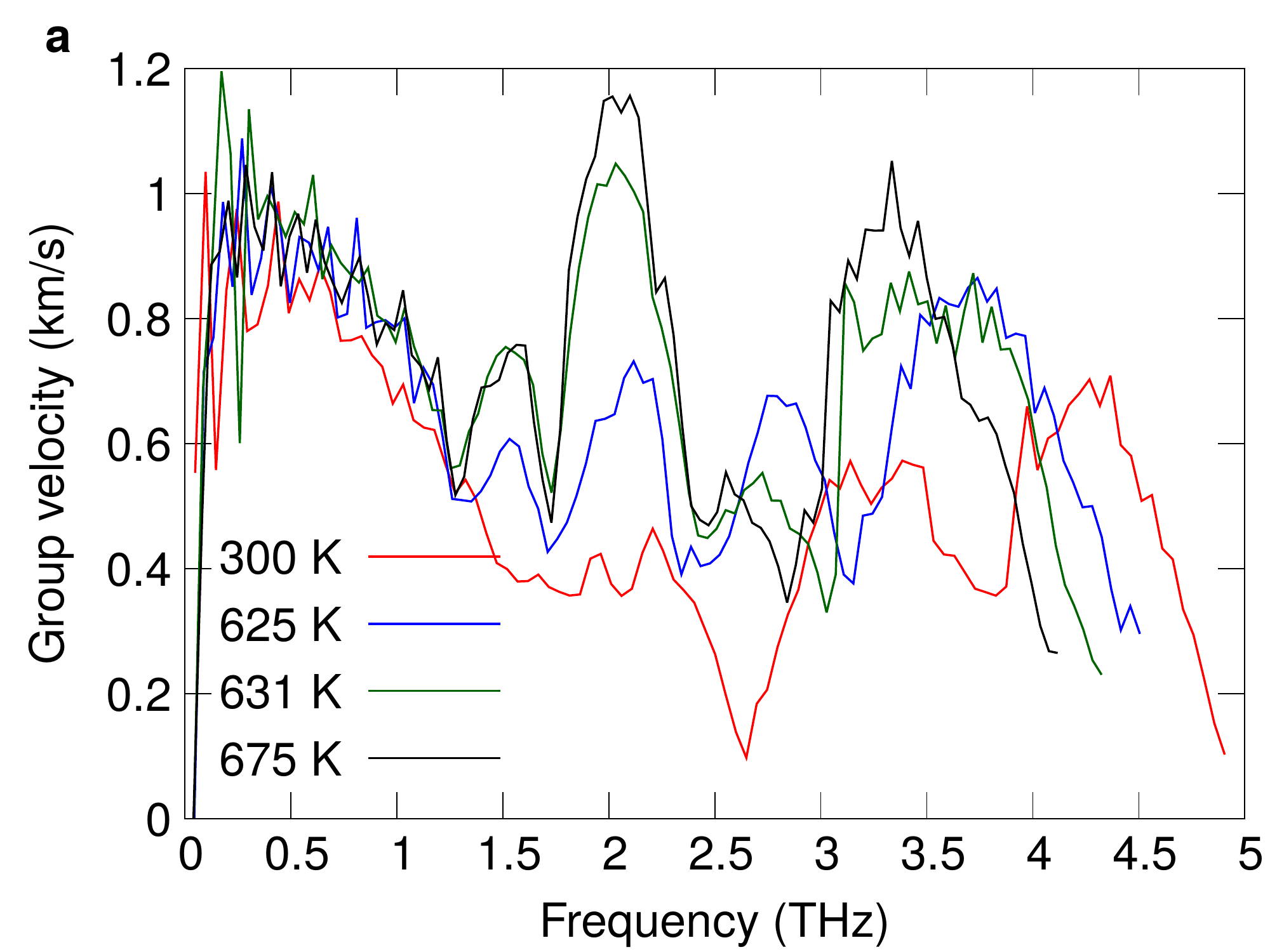}
\includegraphics[width=0.45\textwidth]{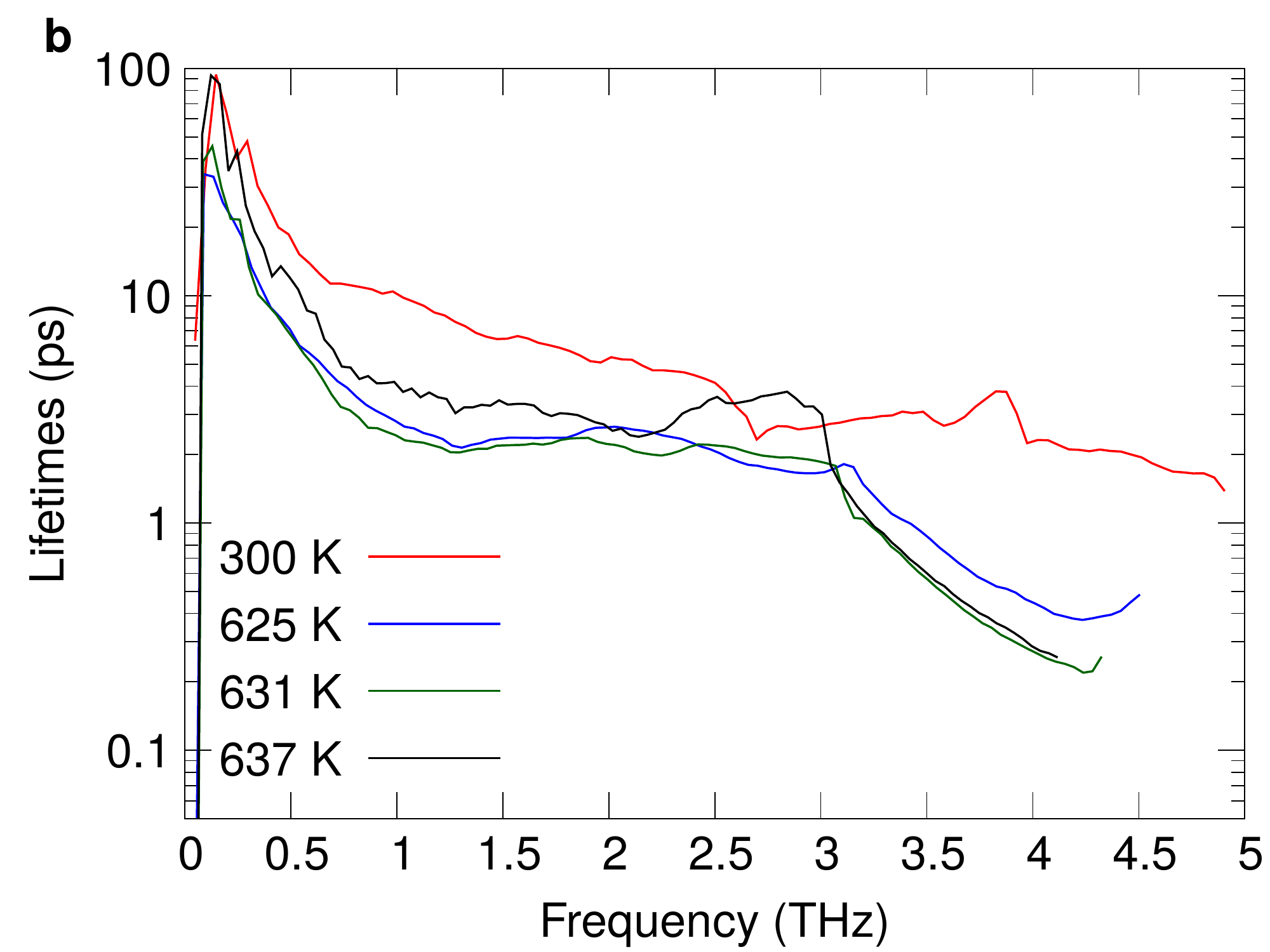}
\caption{\textbf{Frequency dependence of phonon transport properties.} \textbf{a} Average phonon group velocities and \textbf{b} average phonon lifetimes of GeTe versus phonon frequency for different temperatures. Averaging is carried out by convolving the calculated values of these quantities with a Gaussian
 %of specific width
(see Eq.~\ref{eq_av_lifetimes}).}
\label{fig5}
\end{center}
\end{figure}

The phonon group velocities of GeTe are mostly independent of temperature, except very close to the phase transition, see Fig.~\ref{fig5} \textbf{a}. In this temperature region ($600-675$ K), there is an increase in the phonon group velocities across most of the frequency range and most noticeably for phonons between 1 and 3 THz. This is the frequency region that contributes most to the thermal conductivity (see Supplementary Note 4). We thus conclude that the anomalous increase of the thermal conductivity at the phase transition is partially due to this rise in the phonon group velocities. We find that the increase in group velocities originates from the lattice contraction near the critical temperature \cite{Our, main, newmain}. This can be understood from the observation that the increase of phonon group velocities happens most prominently along out of the plane direction, where negative thermal expansion occurs \cite{Our}. 

Phonon lifetimes in weakly anharmonic materials usually follow the 1/T law, similar to thermal conductivity. This is the case in our calculations in the rhombohedral phase far from the phase transition. At the phase transition, however, the phonon lifetimes of acoustic modes decrease more than expected from the 1/T scaling. This is a signature of stronger anharmonicity of acoustic phonon modes closer to the phase transition in the rhombohedral phase. Optical modes have more complicated behaviour. While soft optical modes near the zone centre have much lower lifetimes at the phase transition, this is not the case for the two lowest optical modes in the rest of the Brillouin zone. Even more intriguing is the behaviour of the highest frequency optical modes. At low temperatures, they usually have frequency independent lifetimes, but at the phase transition they decrease exponentially with frequency. The calculated phonon linewidths of the zone centre phonon modes are the same order of magnitude as the measured ones \cite{SoftmodeGeTe} (see Supplementary Note 5).

In the cubic phase there is a substantial increase in the phonon lifetimes with respect to the rhombohedral phase (see Fig.~\ref{fig5} \textbf{b}). The phonon lifetimes at 637 K are larger in most of the frequency range compared to the temperatures closest to the phase transition in the rhombohedral phase (625 K and 631 K). Interestingly, the phonon lifetimes at 675 K are larger than the phonon lifetimes at 300 K rescaled by temperature (i.e. by 675 K/300 K), revealing stronger intrinsic anharmonicity of the rhombohedral phase. Third order force constants are much stronger in the rhombohedral phase, even for very similar temperatures and structures (631 K vs 637 K), see Supplementary Note 6. 

In conclusion, the increase of the lattice thermal conductivity near the phase transition can be attributed to two phenomena, depending on the structure of GeTe. In the rhombohedral phase, the increase in the thermal conductivity is due to negative thermal expansion that causes phonon group velocities to increase, increasing phonon mean free paths. On the other hand, in the cubic phase, phonon lifetimes increase dramatically due to lower intrinsic anharmonicity of this high-symmetry phase. 

\textbf{Lattice thermal conductivity using the Green-Kubo method.} The non-Lorentzian behaviour of the phonon spectral function raises the question whether the Boltzmann transport equation employed in the calculation of the $\kappa$ values in Fig.~\ref{fig4} is valid close to the phase transition. It is assumed in the derivation of the Boltzmann equation that phonons are well defined quasiparticles with unique frequencies and lifetimes. This implies that their spectral weights are the Lorentzian functions centred at the harmonic or TDEP frequencies and the widths equal to the phonon lifetimes. However, evidently this does not hold close to the phase transition, see Fig.~\ref{fig3}.

We include the effect of non-Lorentzian lineshapes on lattice thermal conductivity following the Green-Kubo approach derived in Refs. \cite{srivastava, Sharma}. In this approach, the heat current in the Cartesian direction $j$ is defined as \cite{srivastava, Hardy, Krumhansl}:
\begin{align*}
J^{j} (t)= \frac{1}{2NV}\sum _{\vec{q}, s,s'}\hbar\omega _{\vec{q}, s} p_{\vec{q},s}(t) u ^{\dagger}_{\vec{q},s}(t) v ^{j} _{\vec{q},s},  \numberthis
\end{align*}    
where $v^{j} _{\vec{q},s}$ is the group velocity of the phonon mode with wave vector $\vec{q}$ and branch $s$, $V$ is the volume of the unit cell, $N$ is the number of $\vec{q}$ points, and $p_{\vec{q},s} (t)$ and $u_{\vec{q},s} (t)$ are the Fourier transforms of the phonon momentum and position operators. The lattice thermal conductivity tensor is obtained by employing the Green-Kubo relation \cite{GreenKubo}:
\begin{align*}
\kappa ^{i,j} = \frac{NV}{k_{B}T^2} \Re \int \left\langle J^{i}(0)J^{j}(t)\right\rangle \text{d}t. \numberthis
\end{align*}
The spectral theorem is then used to relate the heat current autocorrelation function in the equation above to the one particle retarded Green function \cite{Maradudin, Pathak, Behera, Sharma, Zubarev}. The final expression for the thermal conductivity in this approach is \cite{srivastava, Sharma}:
\begin{align*}
 & \kappa ^{i, j} = 4 \frac{\hbar ^2 k_{B}\beta ^2}{N V \pi} \sum _{\vec{q}, s, s'} \omega ^{2}_{\vec{q}, s}(\omega ^{2}_{\vec{q}, s} + \omega ^{2}_{\vec{q}, s'})\times  \\
& v ^{i}_{\vec{q},s,s'}v ^{j}_{\vec{q},s,s'} \int _{-\infty} ^{\infty} \text{d}\Omega \frac{\exp{(\beta\hbar\Omega})}{\left(\exp{(\beta\hbar\Omega)} - 1\right)^2}\times  \\
	&\frac{\Omega ^2 \Gamma _{\vec{q}, s}(\Omega)\Gamma _{\vec{q}, s'}(\Omega)}{\left[ \epsilon ^2 _{\vec{q}, s} + 4\omega_{\vec{q},s}^2\Gamma ^2 _{\vec{q}, s}(\Omega)\right]\left[ \epsilon ^2 _{\vec{q}, s'} + 4\omega_{\vec{q},s'}^2\Gamma ^2 _{\vec{q}, s'}(\Omega)\right]}, \numberthis
\label{eq4} 
\end{align*}
where $\epsilon _{\vec{q},s}$ is:
\begin{align*}
\epsilon _{\vec{q}, s} = \Omega ^2 - \omega ^2 _{\vec{q},s} - 2\omega _{\vec{q},s}\Delta _{\vec{q},s} (\Omega). \numberthis
\end{align*}
$\vec{v}_{\vec{q},s,s'}$ is the generalized phonon group velocity \cite{Hardy}:
\small
\begin{align*}
\vec{v}_{\vec{q},s,s'} = \frac{i}{2\sqrt{\omega _{\vec{q},s}\omega _{\vec{q},s'}}}\sum _{a,b} X ^{a} _{\vec{q},s}\sum _{\vec{R}}\left( \vec{R}\frac{\Phi _{a,b}(\vec{R})}{\sqrt{m_{a}m_{b}}} \right)X ^{b} _{\vec{q},s'}, \numberthis
\end{align*} 
\normalsize
where $\vec{X} _{\vec{q},s}$ is the eigenvector of the phonon with wave vector $\vec{q}$ and branch $s$, $\Phi _{a,b}(\vec{R})$ is the force constant between atoms with masses $m_{a}$ and $m_{b}$ and the vector distance $\vec{R}$. 

 We can separate Eq.~\ref{eq4} into two parts. The first part is diagonal ($s = s'$). In the limit of small anharmonicity ($\Delta _{\vec{q},s} = 0$ and $\Gamma _{\vec{q},s} \ll \omega _{\vec{q},s}$), this part reduces to the standard solution of the Boltzmann equation in the relaxation time approximation. The second, non-diagonal part ($s \neq s'$), can be reduced in the limit of small anharmonicity to the expressions similar to the ones given in Refs.~\cite{Simoncelli, Baroni}. The non-diagonal contribution to the lattice thermal conductivity will become prominent only if there is a substantial overlap in the spectral functions of two phonon modes with the same wave vector. This is only true in the case of strong anharmonicity or when spectral functions broaden due to disorder. 

\begin{figure}[h!]
\begin{center}
\includegraphics[width=0.45\textwidth]{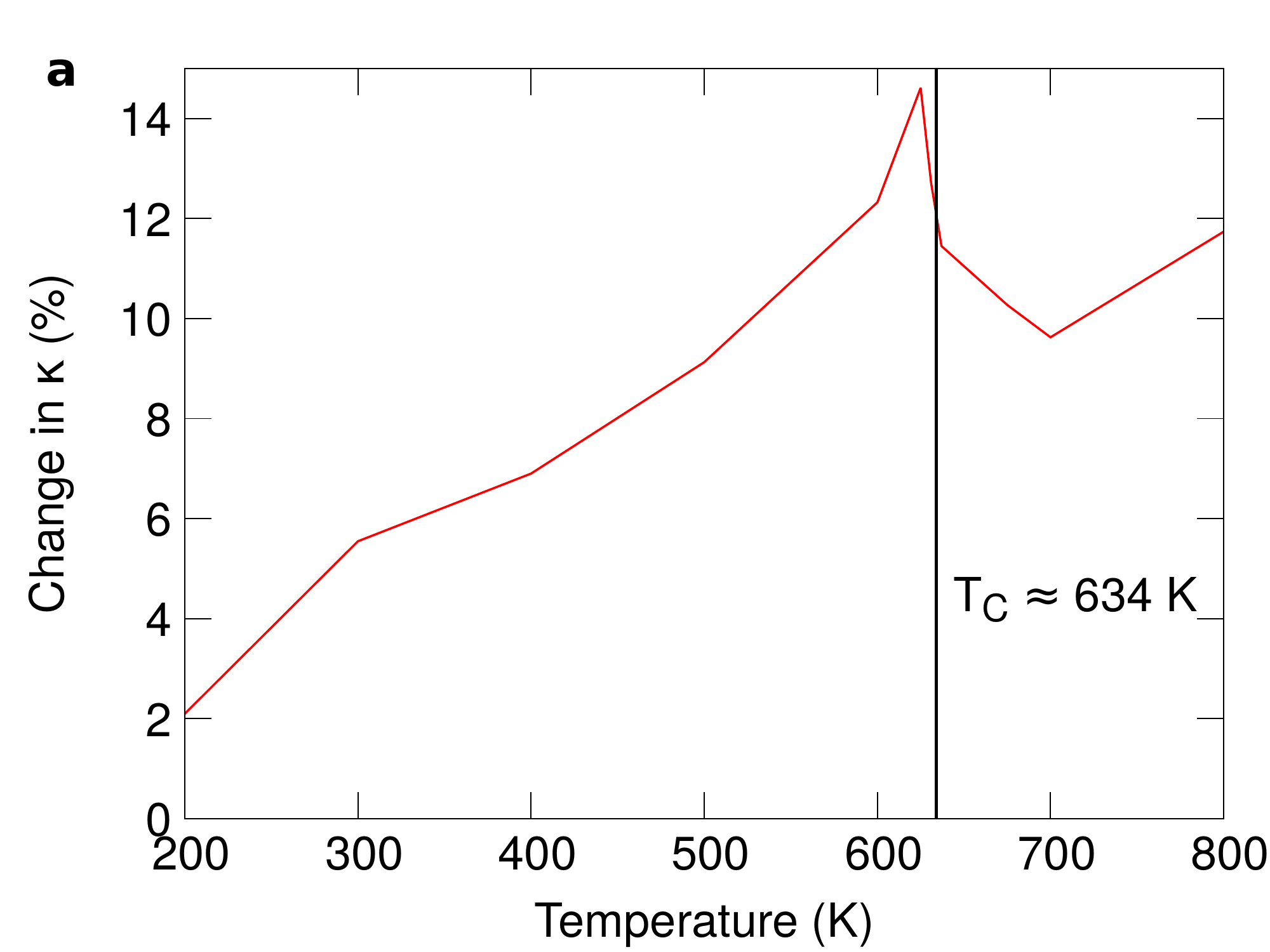}
\includegraphics[width=0.45\textwidth]{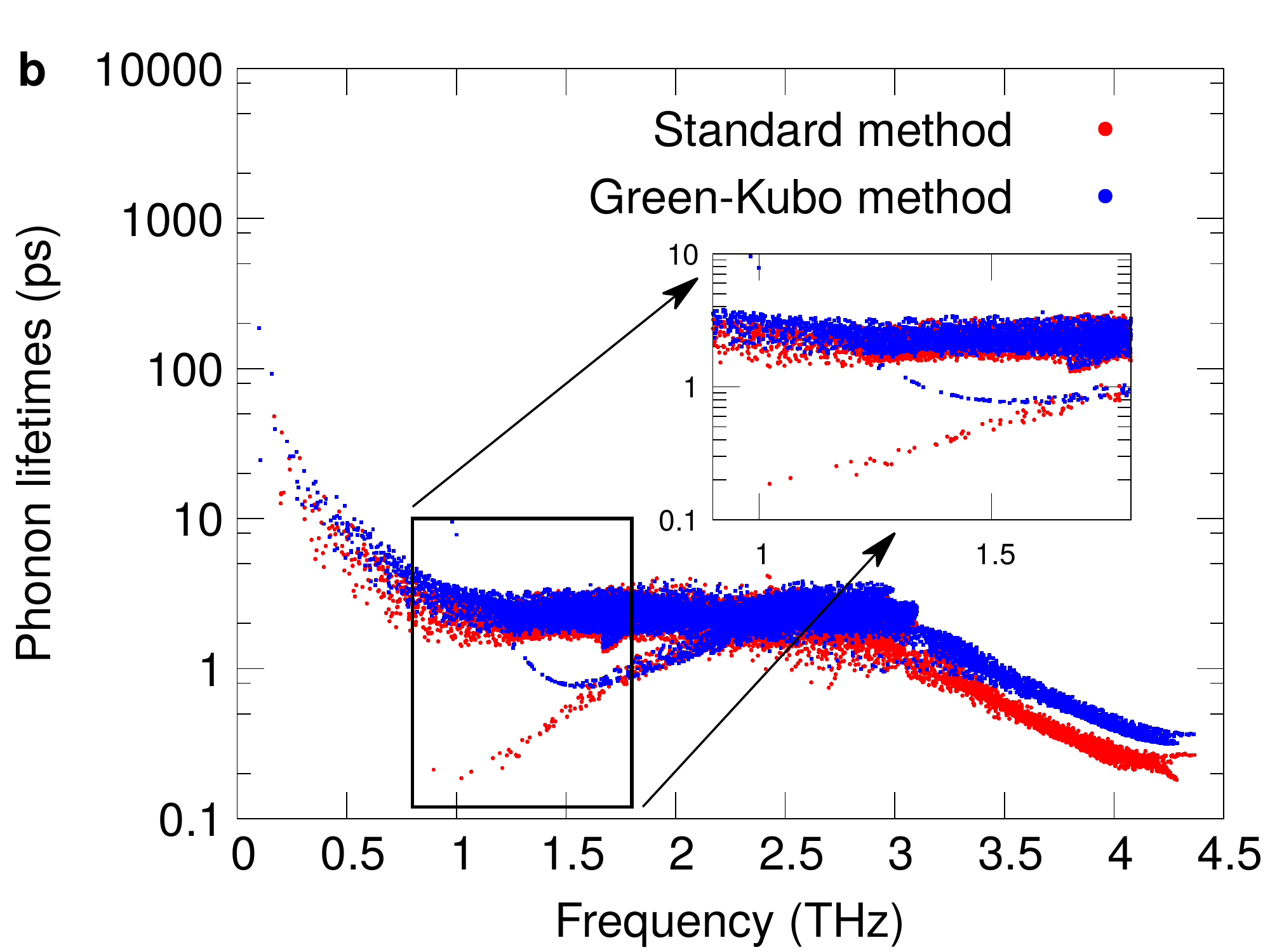}
\caption{\textbf{Green-Kubo lattice thermal conductivity.} \textbf{a} Difference in the calculated lattice thermal conductivity using the Green-Kubo method (Eq.~\ref{eq4}) and the Boltzmann transport equation. \textbf{b} Phonon lifetimes of GeTe (see text for explanation) in the Green-Kubo and Boltzmann transport equation approach plotted versus TDEP frequency of phonon modes. Inset shows the region of soft phonon modes where the change is most noticeable.}
\label{fig6}
\end{center}
\end{figure}

We have implemented the expression given by Eq.~\ref{eq4} with the TDEP method. Figure~\ref{fig6} \textbf{a} shows the difference between our results obtained using Eq.~\ref{eq4} and BTE (Fig. ~\ref{fig4}). We can see that the BTE underestimates the thermal conductivity in the whole temperature range. Additionally, we can see that the underestimation is not large, around 10\% even at high temperatures. Overall, the difference scales linearly with temperature. We can also see that the difference is largest at the phase transition, which is expected considering large deviations from the Lorentzian shape of the phonon spectral functions in this region (see Fig.~\ref{fig2}). The contribution of the non-diagonal part of the lattice thermal conductivity is comparable to the overall enhancement of the diagonal part due to non-Lorentzian shapes of the phonon spectral functions.

To understand the reason for the increased difference in $\kappa$ at the phase transition obtained by the standard BTE method and using the Green-Kubo relation (Eq.~\ref{eq4}), we show the phonon lifetimes of GeTe calculated using the two methods in Fig.~\ref{fig6} \textbf{b}. We define the phonon lifetimes in the Green-Kubo method as:
\begin{align*}
\tau ^{GK}_{\vec{q},s} = 8\frac{\hbar ^2 k_{B}\beta ^2 \omega ^4 _{\vec{q},s}}{\pi c_{\vec{q},s}} & \int _{-\infty} ^{\infty} \text{d}\Omega \frac{\exp{(\beta\hbar\Omega})}{\left(\exp{(\beta\hbar\Omega)} - 1\right)^2}\times \numberthis \\
	&\frac{\Omega ^2 \Gamma _{\vec{q}, s}^2(\Omega)}{\left[ \epsilon ^2 _{\vec{q}, s} + 4\omega_{\vec{q},s}^2\Gamma ^2 _{\vec{q}, s}(\Omega)\right]^2}, 
\end{align*}
where $c_{\vec{q},s}$ is the harmonic heat capacity of the phonon mode ($\vec{q},s$). The increase of the phonon lifetimes in the Green-Kubo method is visible in the whole Brillouin zone. It is, however, most prominent in the region of soft phonon modes, where the phonon lifetimes increase by a factor of 100. The increase in phonon lifetimes mostly comes from the shifts of the peaks of the spectral functions, which increases the heat capacity of the phonon modes compared to the harmonic heat capacity.   

\textbf{Discussion.}
%We can apply the same method to evaluate
Here we highlight the advantages of the Green-Kubo method we implemented here over the standard approaches for computing $\kappa$ in strongly anharmonic materials. Unlike the Boltzmann transport equation, the Green-Kubo method accounts for non-Lorentzian shapes of phonon spectral functions. It is possible to include these effects also by running long molecular dynamics (MD) simulations. Compared to MD, the Green-Kubo method presented here is faster and easier to converge, which is particularly important if these methods are combined with first principles calculations. The results obtained using this approach are easier to interpret and analyse compared to traditional MD approaches. Unlike MD simulations, the Green-Kubo method uses the Bose-Einstein statistics for phonons. Additionally, this method gives the non-diagonal part of lattice thermal conductivity, accounting explicitly for the whole phonon power spectra. The method of Refs.~\cite{Simoncelli, Baroni} includes only the values of the phonon self-energy at the harmonic frequency in the evaluation of the non-diagonal contribution to $\kappa$ and is not applicable to strongly anharmonic materials. 

In conclusion, we have performed a detailed first principles study of the lattice thermal conductivity $\kappa$ of GeTe close to the ferroelectric phase transition. The temperature dependent effective potential (TDEP) frequencies of the soft modes, although dramatically softened, do not become zero at the phase transition. On the other hand, strong anharmonicity causes the spectral functions of the soft modes to collapse and effectively peak at zero frequency. Strong anharmonicity  minimizes the acoustic phonon modes lifetimes at the phase transition. However, we calculate an increase in the lattice thermal conductivity at the phase transition, in agreement with experiments. In the rhombohedral phase, this effect is due to negative thermal expansion that increases phonon group velocities. In the cubic phase, the increase in $\kappa$ is primarily driven by increased phonon lifetimes due to smaller anharmonicity of phonon modes compared to the rhombohedral phase. We implement a novel approach to compute lattice thermal conductivity that includes the observed non-Lorentzian power spectra of phonon modes. Using the new approach, we find that the calculated $\kappa$ increases even further at the phase transition, which is the consequence of larger phonon populations due to softening of the phonon modes caused by phonon-phonon interaction. 

\section{Computational methods.}
We calculate the lattice thermal conductivity of GeTe from first principles using density functional theory (DFT) and the Boltzmann transport equation (BTE). In this approach, the lattice thermal conductivity tensor is given as \cite{srivastava}:
\begin{align*}
\kappa ^{i,j} = \frac{1}{NV}\sum _{\vec{q}, s} c_{\vec{q}, s} v^{i} _{\vec{q}, s} v^{j} _{\vec{q}, s} \tau _{\vec{q}, s}, \numberthis
\end{align*}
where $i$ and $j$ are the Cartesian directions, $N$ is the number of $\vec{q}$ points, $V$ is the unit cell volume, $c _{\vec{q}, s}$ is the phonon mode heat capacity and $v^{i} _{\vec{q}, s}$ is the group velocity of the phonon mode $(\vec{q}, s)$ in the direction $i$. The relaxation time of the same mode is $\tau _{\vec{q}, s} = 1/2\Gamma _{\vec{q}, s}$, where the imaginary part of the phonon self-energy due to three-phonon scattering is given as \cite{Maradudin}: 
\small
\begin{align*}
\Gamma _{\lambda} = \frac{\pi\hbar}{16N}\sum _{\lambda{'},\lambda{''}} & |\Phi _{\lambda \lambda{'} \lambda{''}}|^2 \{ (n_{\lambda{'}} + n_{\lambda{''}} + 1)\delta (\omega _{\lambda} - \omega _{\lambda{'}} - \omega _{\lambda{''}}) \\ & +
 2(n_{\lambda{'}} - n_{\lambda{''}})\delta (\omega _{\lambda} - \omega _{\lambda{'}} + \omega _{\lambda{''}}) \} . \numberthis
\end{align*}
\normalsize
Here $\lambda$ is a short hand notation for $(\vec{q}, s)$ and phonon momentum is conserved in the three-phonon processes above. The three phonon matrix element $\Phi _{\lambda \lambda{'} \lambda{''}}$ is calculated using:
\begin{align*}
\Phi_{\lambda\lambda'\lambda''} =\sum_{ijk}\sum_{\alpha\beta\gamma}& \frac{X_{\lambda}^{i \alpha}X_{\lambda'}^{j \beta}X_{\lambda''}^{k \gamma}}{\sqrt{m_{i}m_{j}m_{k}}\sqrt{\omega_{\lambda}\omega_{\lambda'}\omega_{\lambda''}} } \times \\ &
\times \Phi^{\alpha\beta\gamma}_{ijk} e^{i \vec{q}\cdot\vec{r}_i + i \vec{q}'\cdot\vec{r}_j+i \vec{q}''\cdot\vec{r}_k}, \numberthis
\end{align*}
where $m _{i}$ is the mass of atom $i$, $X_{\lambda}^{i \alpha}$ is the component $\alpha$ of the eigenvector for mode $\lambda$ and atom $i$, and $\vec{r}_i$ is the vector associated with atom $i$. $\omega _{\lambda}$ is the TDEP frequency of the phonon mode $\lambda$, and $\Phi^{\alpha\beta\gamma}_{ijk}$ is the third order interatomic force constant. This expression for the imaginary part of the self-energy is the result of the first order perturbation theory for the phonon self-energy due to third-order anharmonicity (so called bubble term in the diagrammatic representation of the self energy). The real part of self-energy is the Kramers-Kronig transformation of the imaginary part. 

The temperature evolution of the crystal lattice of GeTe was calculated using molecular dynamics (MD). We ran MD calculations using the LAMMPS software~\cite{lammps}. To perform MD simulations, we developed a very accurate interatomic potential based on the Gaussian Approximation Potentials scheme \cite{GAP1, GAP2} (see Supplementary Note 7). To obtain structural parameters at the temperature $T$, we ran the NPT ensemble MD calculation of a 2000 atom cell (the $10\times 10\times 10$ supercell) fixing pressure to 0 Pa (we checked the convergence with respect to simulation cell size and found the 2000 atom cell to be converged). Following a 10 and 20 ps equilibration in the NVT and NPT ensembles respectively, we ran simulation for 100 ps sampling the lattice constant $a$, the rhombohedral angle $\theta$ and the volume every 100 fs. For the calculation of the interatomic displacement parameter $\mu$, we sampled atomic positions every timestep (1 fs). Final structural parameters were obtained as a simple sample average and are given in Supplementary Note 8. We obtain negative thermal expansion at the phase transition~\cite{main, mdpigete, GeTeBo, Our}. In our MD simulations, the system is still ferroelectric at 631 K, while it is paraelectric at 637 K. Because of this, we define the critical temperature as the average of these two temperatures (T$_{\text{C}}$ = 634 K).

To get temperature dependent interatomic force constants, we use the temperature dependent effective potential (TDEP) method \cite{TDEP1, TDEP2, TDEP3}. This approach employs a fitting procedure of the second and third order force constants to DFT forces on forces sampled along a molecular dynamics (MD) trajectory. We ran MD simulations using the GAP potential in the NVT ensemble on the structures obtained from the MD study of thermal expansion to obtain atomic configurations. After 50 ps of equilibration we sampled 24 configurations on the 300 ps long trajectory. We used a 512 atom supercell to converge phonon properties with respect to the second and third order force constants cutoff (12  and 8 \r{A}, respectively). We extracted selected configurations and carried out density functional theory calculations on them to obtain forces. Since the 0 K structures of GeTe obtained using GAP and DFT are slightly different (see Supplementary Note 7), we obtained the DFT structure of GeTe at a certain temperature using the thermal expansion coefficients calculated with MD combined with the 0 K DFT structure.

We note that we did not include LO/TO splitting due to long-range electrostatic forces in any of our calculations. GeTe has a large number of intrinsic vacancies which give large free charge carrier concentrations. Free charge carriers will perfectly screen the long-range interaction~\cite{SoftmodeGeTe}, resulting in no LO/TO splitting in real samples.

DFT calculations were performed using the ABINIT software package~\cite{ABINIT, ABINIT2}.~We use a generalized gradient approximation with the Perdew-Burke-Ernzerhof parametrization (GGA-PBE) \cite{GGAPBE} for the exchange-correlation functional and the Hartwigsen-Goedecker-Hutter (HGH) pseudopotentials \cite{HGHpseudo}. Wave functions are represented in a plane wave basis set with the cutoff of 16 Ha, and the $\Gamma$ point is used for sampling of electronic states. Sampling of phonon states is carried out using a $30\times30\times30$ $\vec{q}$-point grid.

\section{Code availability.}
The code that implements the temperature effective potential (TDEP) method is available from Olle Hellman upon reasonable request. The additional data processing scripts and codes are available from the corresponding authors. GAP software is available for non-commercial use from www.libatoms.org.

\section{Data availability.}
The authors declare that the data supporting the present work is available from the corresponding authors upon reasonable request.

\section{Acknowledgements.}
This work is supported by Science Foundation Ireland under grant numbers 15/IA/3160 and 13/RC/2077. The later grant is cofunded under the European Regional Development Fund. O.H. gratefully acknowledges support from the Knut and Alice Wallenberg Foundation (Wallenberg Scholar Grant No. KAW-2018.0194) and the Swedish Government Strategic Research Areas SeRC. We acknowledge the Irish Centre for High-End Computing (ICHEC) for the provision of computational facilities.

\section{Author contributions.}
I.S. and {\DJ}.D. conceived the research plan. {\DJ}.D. performed the calculations. I.S. supervised the work. {\DJ}.D. and I.S. wrote the manuscript with contribution from all authors. O.H. provided the code for the TDEP method. All authors discussed and interpreted the results.

\section{Competing interests.} 
The authors declare no competing financial or non-financial interests.

\section*{References.}

\bibliographystyle{naturemag}
\bibliography{main}{}

\newpage

\section{Supplementary note 1: Phonon frequencies at the phase transition}

We first justify why one should compare "anharmonic" frequencies with experiment. Experiments can not directly measure phonon frequencies. What experiments actually measure (for example neutron scattering experiments) are the scattering cross sections. In the example of the neutron scattering experiment, the lowest contribution to the scattering cross section is the displacement autocorrelation function ~\cite{Maradudin} (we expect the similar expression for Raman scattering, the difficulty is whether the polarizability operator is linearly dependent on the displacement operator \cite{Cowley_Raman}). We can directly calculate this quantity (subject to some approximations obviously) using second and third order force constants. In the experiment one recognizes a peak of the signal (the scattering cross section for the incident neutron) in the energy space as the phonon frequency. We do the exact same thing when we determine the anharmonic frequency. We calculate the displacement autocorrelation function, find at which frequency it peaks and assign that frequency as the anharmonic frequency.

Now let us discuss the nature of the phase transition in a model system described by a double well potential as defined in Ref.~\cite{landau}. This will help us explain why we expect the temperature dependent effective potential (TDEP) frequency to fall to zero at the displacive phase transition. The total energy of this system ($\Phi$) is given by:
\begin{align}
\Phi= \Phi _{0} + Ap^2 + Bp^4 , 
\label{landeq}
\end{align}
where we dropped the explicit dependence of $A$ and $B$ on thermodynamic quantities and the order parameter is denoted as $p$. Ref.~\cite{landau} assumes $A$ and $B$ are temperature and pressure dependent. With that in mind one can interpret Eq.~\ref{landeq} as the best possible fit of the true energy to the functional form expressed above. This reasoning is identical to the TDEP method.

The important detail is that $p$ is the order parameter. In the ferroelectric system, the order parameter is the polarization ($p = \frac{1}{\Delta V}\sum_{\Delta V} p_i$; we sum local dipole moments in some volume $\Delta V$). In the lowest approximation, the local (unit cell) polarization $p_i$ is directly proportional to the interatomic displacement parameter (the definition of Born effective charge). In the displacive phase transition, the polarization and the interatomic displacement parameter $\mu$ are the same ($p \equiv \mu$), since all of the local dipole moments point in the same direction ($\mu$ are all in the same minimum of the double well). However, in the order-disorder phase transition, the polarization and the interatomic displacement parameter are not the same. The interatomic displacement parameter shows the local minimum of the energy (a degenerate quantity), while the polarization is the average over them. If the interatomic displacement parameters are unformly distributed between two or more minima, the polarization is likely to be zero, while $\mu$ is not.

In the low symmetry phase, the parameter $A$ in Eq.~\ref{landeq} is negative. In the high symmetry phase, it is positive. The phase transition is defined as a point where this parameter changes sign ~\cite{landau}. If the phase transition is displacive, and the polarization is the same as $\mu$, one can recognize that $A$ is proportional to the square of the phonon frequency of the high-symmetry phase (for $A$ to be related to a phonon frequency, $p$ must be equal to the small perturbation from the equilibrium position, and if the system is in the low symmetry phase, $\mu$ is not the perturbation from the equilibrium position, it is the equilibrium position). This frequency can be interpreted as the TDEP frequency due to the similar nature of $A$ and TDEP force constants. $A$ has to pass through zero at the phase transition (both displacive and order disorder), meaning that in the displacive phase transition the TDEP frequency must pass through zero as well. In the order disorder phase transition, however, the order parameter (the polarization $p$) is not the same as $\mu$, it is actually the average of all local $\mu$-s. This means that we can not associate $A$ with the TDEP phonon frequency of the high symmetry phase. The parameter $A$ still has to become zero at the phase transition, but it is not related to the TDEP phonon frequency. In the order-disorder phase transition, we can not make any claims, whether the TDEP phonon frequency goes to zero or not. On the other hand, the connection between the "anharmonic" frequency and the parameter $A$ is not obvious at all.

Finally, we note that scattering experiments used to measure phonon frequencies do not measure TDEP frequencies (they measure the peak of the power spectrum), so they can not conclusively tell if the phase transition is of the displacive type or not. 

\section{Supplementary note 2: Phonon spectral function of GeTe in the cubic phase}

\begin{figure}
\begin{center}
\includegraphics[width=0.45\textwidth]{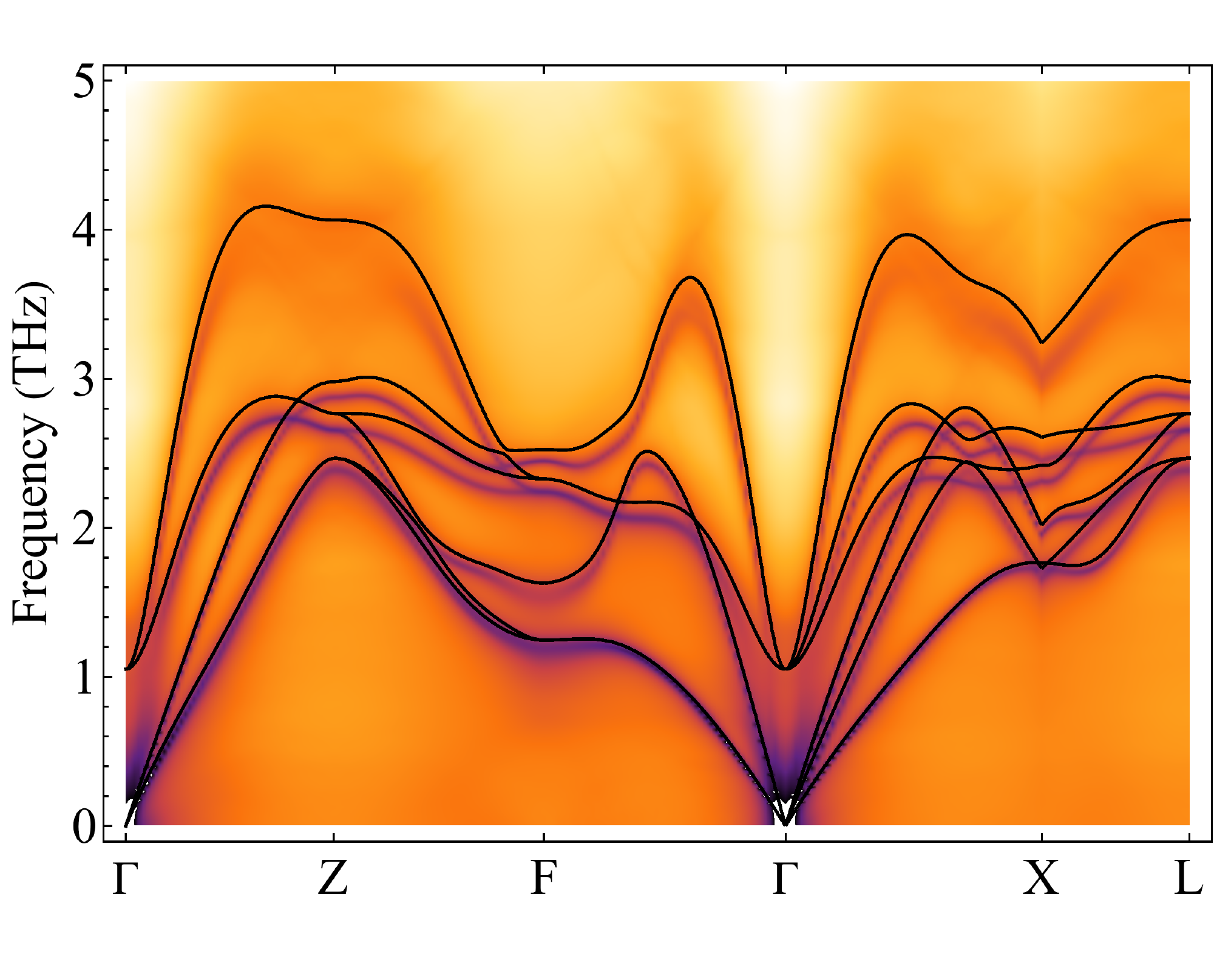}
\caption{Phonon spectral function of GeTe in the cubic phase at 637 K. The black lines are the TDEP frequencies calculated at this temperature.}
\label{supfig8}
\end{center}
\end{figure}

Supplementary figure~\ref{supfig8} shows the phonon spectral function of GeTe in the cubic phase at 637 K. The black lines are the calculated TDEP frequencies of GeTe at this temperature. We can see the similar behavior of the phonon spectral function as at 631 K (see Fig. 2 of the main manuscript). The zone center optical modes can not be resolved in the phonon spectral function. 

\section{Supplementary note 3: Spectral function of the zone centre optical modes near the phase transition}

\begin{figure}
\begin{center}
\includegraphics[width=0.45\textwidth]{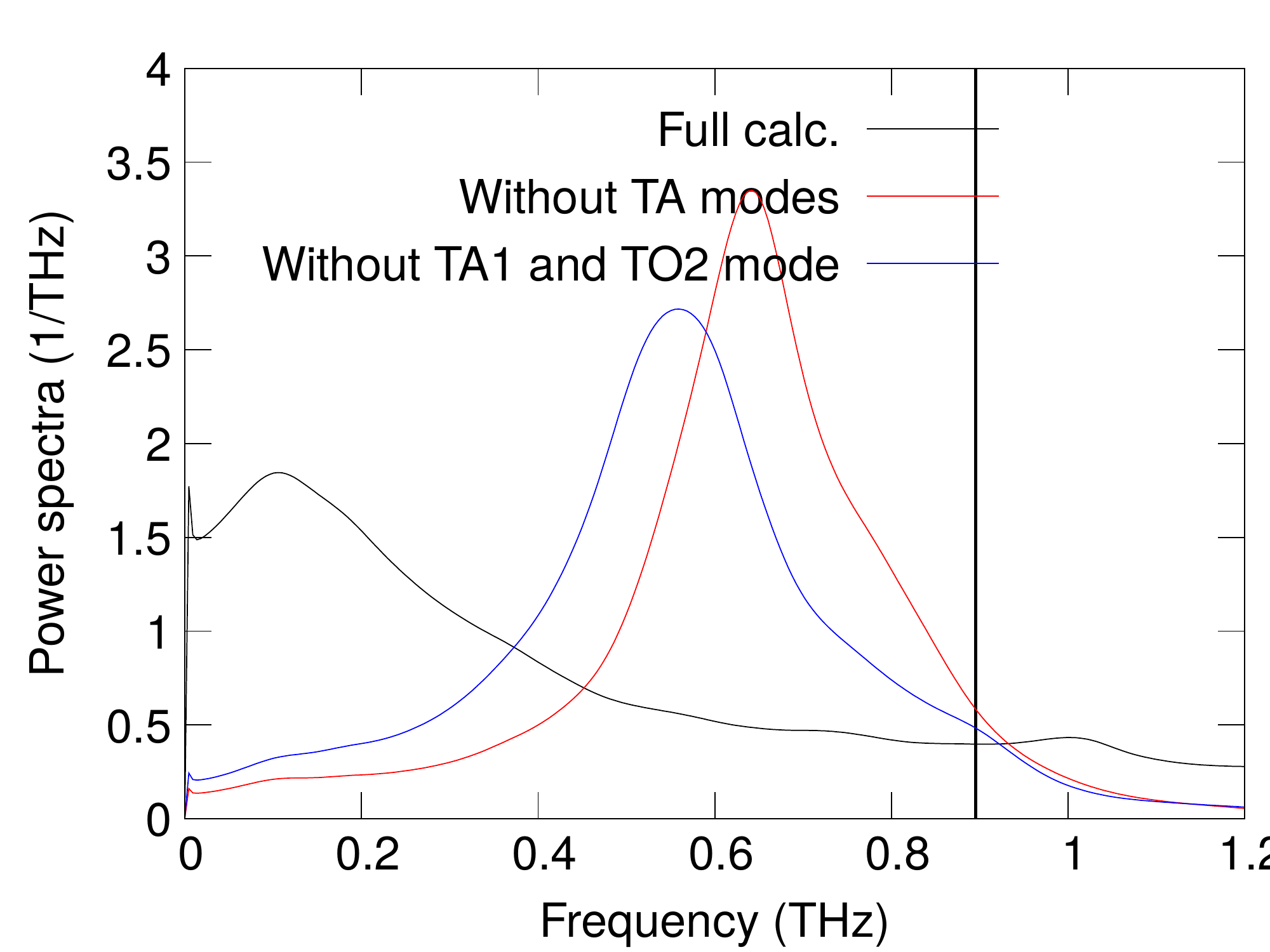}
\caption{Spectral function of the zone centre A$_1$ mode at 631 K including all phonon-phonon interactions (``Full calc.''), excluding the coupling to transverse acoustic (TA) modes (``Without TA modes''), and excluding the interaction with the first transverse acoustic mode and the second transverse optical mode (``Without TA1 and TO2 mode''). The vertical line represents the TDEP frequency of the A$_1$ ($\Gamma$) mode at 631 K.}
\label{supfig1}
\end{center}
\end{figure}

Supplementary figure~\ref{supfig1} shows the spectral function of the zone centre A$_1$ phonon mode at 631 K, computed including and excluding transverse acoustic-A$_1$ mode coupling. When we neglect the coupling of the A$_1$ mode with two lowest modes in our calculation (which we will call transverse acoustic (TA) modes), the peak of the spectral function shifts closer to the TDEP frequency. A similar behaviour is observed if we disregard the coupling of the A$_1$ zone centre mode with TA1 and the second  (mostly transverse) optical (TO2) mode. This illustrates that although coupling to TA modes is strong, it is not the sole source of the exotic behaviour of the soft A$_1$ phonon power spectrum. In Supplementary figure~\ref{supfig1} we show the types of coupling that bring the phonon power spectrum closest to the Lorentzian shape.

The spectral function of the zone centre E mode in GeTe is illustrated in Supplementary figure~\ref{supfig2} \textbf{a} for several temperatures. We can see a secondary peak in the E spectral function at 625 K, indicating that anharmonicity of this mode is even larger than that of the A$_1$ mode. This is somewhat to be expected since the TDEP frequency of the E mode is lower than that of the A$_1$ mode, leading to larger coupling to the rest of the modes. At 631 K the spectral function of this mode has a peak at almost 0 THz. In the rocksalt phase, the E and A$_1$ modes are degenerate and have the same lineshapes. 

In Supplementary figure~\ref{supfig2} \textbf{b} we show the spectral function of the zone centre E mode at 631 K with full anharmonicity, excluding the coupling of E mode with transverse acoustic modes and excluding the coupling to the TA2 mode and the first (mostly transverse) optical (TO1) mode. We can see that the Lorentzian shape of the spectral function is not regained after excluding different types of interaction and thus we conclude that the non-Lorentzian shape of the E mode power spectrum and the softening of the E frequency is due to coupling to the entire phonon bath, rather than a particular phonon branch. However, here we can say that the coupling to the acoustic modes is dominant, as evidenced by the strongest renormalization of the E mode spectral function.  

\begin{figure}
\begin{center}
\includegraphics[width=0.45\textwidth]{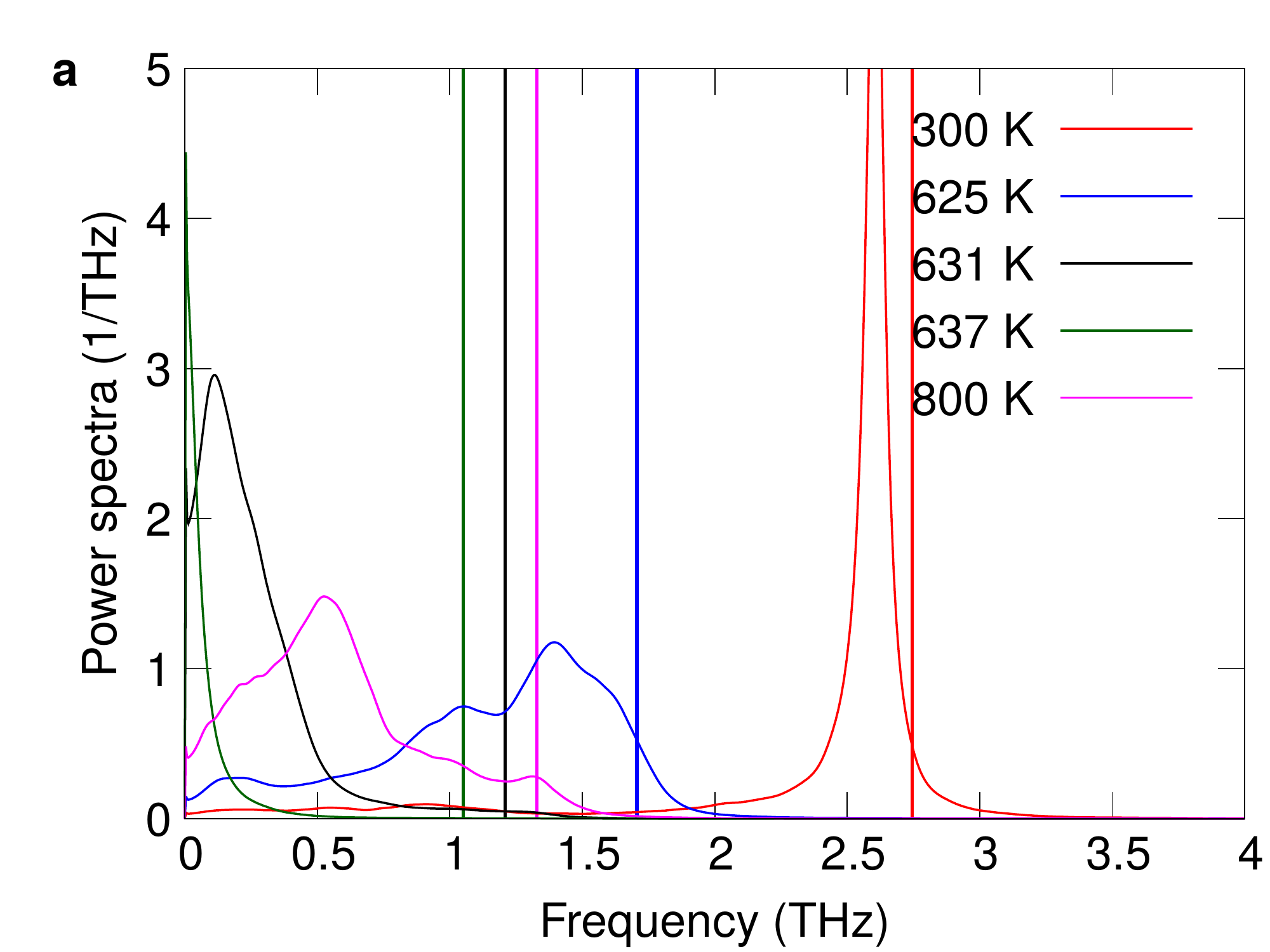}
\includegraphics[width=0.45\textwidth]{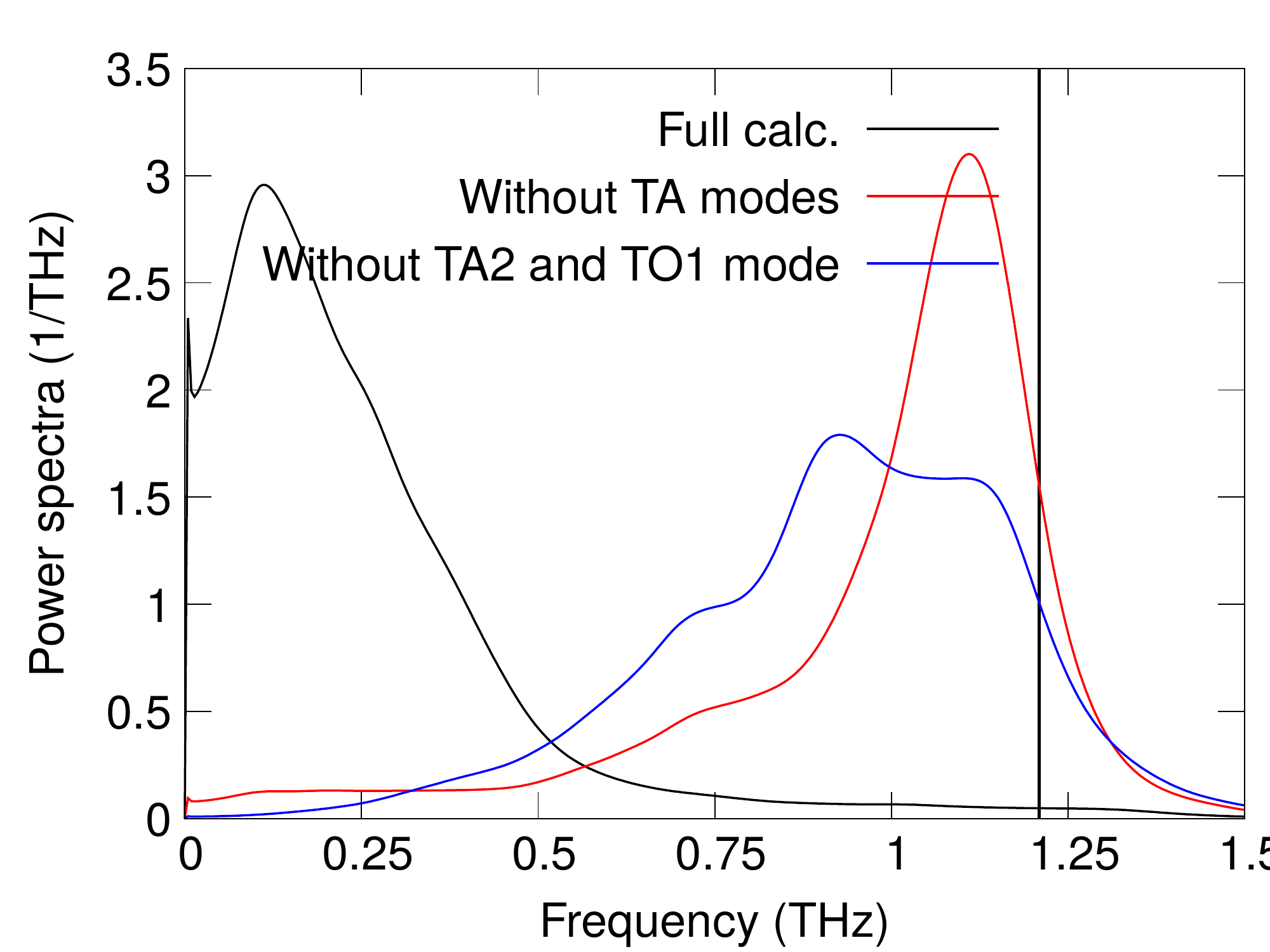}
\caption{\textbf{a} Spectral function of the zone centre E mode at different temperatures. \textbf{b} Spectral function of the E ($\Gamma$) mode at 631 K including all phonon-phonon interactions (``Full calc.''), excluding the coupling to transverse acoustic modes (``Without TA modes'') and excluding the coupling between the second transverse acoustic and first transverse optical modes (``Without TA2 and TO1 modes''). The vertical lines represent the TDEP frequencies of the E ($\Gamma$) mode at various temperatures.}
\label{supfig2}
\end{center}
\end{figure}

\section*{Supplementary Note 4: Spectral thermal conductivity}

To analyze the contribution of specific phonon modes to the total lattice thermal conductivity, we calculated the spectral thermal conductivity at different temperatures by convolving the total lattice thermal conductivity with a Gaussian of an appropriate width (see the main part for more details). We have found that acoustic modes give the dominant contribution to the lattice thermal conductivity, as one would expect (see Supplementary Fig.~\ref{supfig3}). At temperatures near the phase transition, the overall contribution of transverse modes (acoustic and optical) to the lattice thermal conductivity of GeTe diminishes. At the phase transition, the largest contribution to $\kappa$ comes from the phonon modes in the frequency range between 1 and 3 THz. This is the frequency region that shows the most prominent enhancement of phonon group velocities due to negative thermal expansion in the rhombohedral phase.

\begin{figure}
\begin{center}
\includegraphics[width=0.45\textwidth]{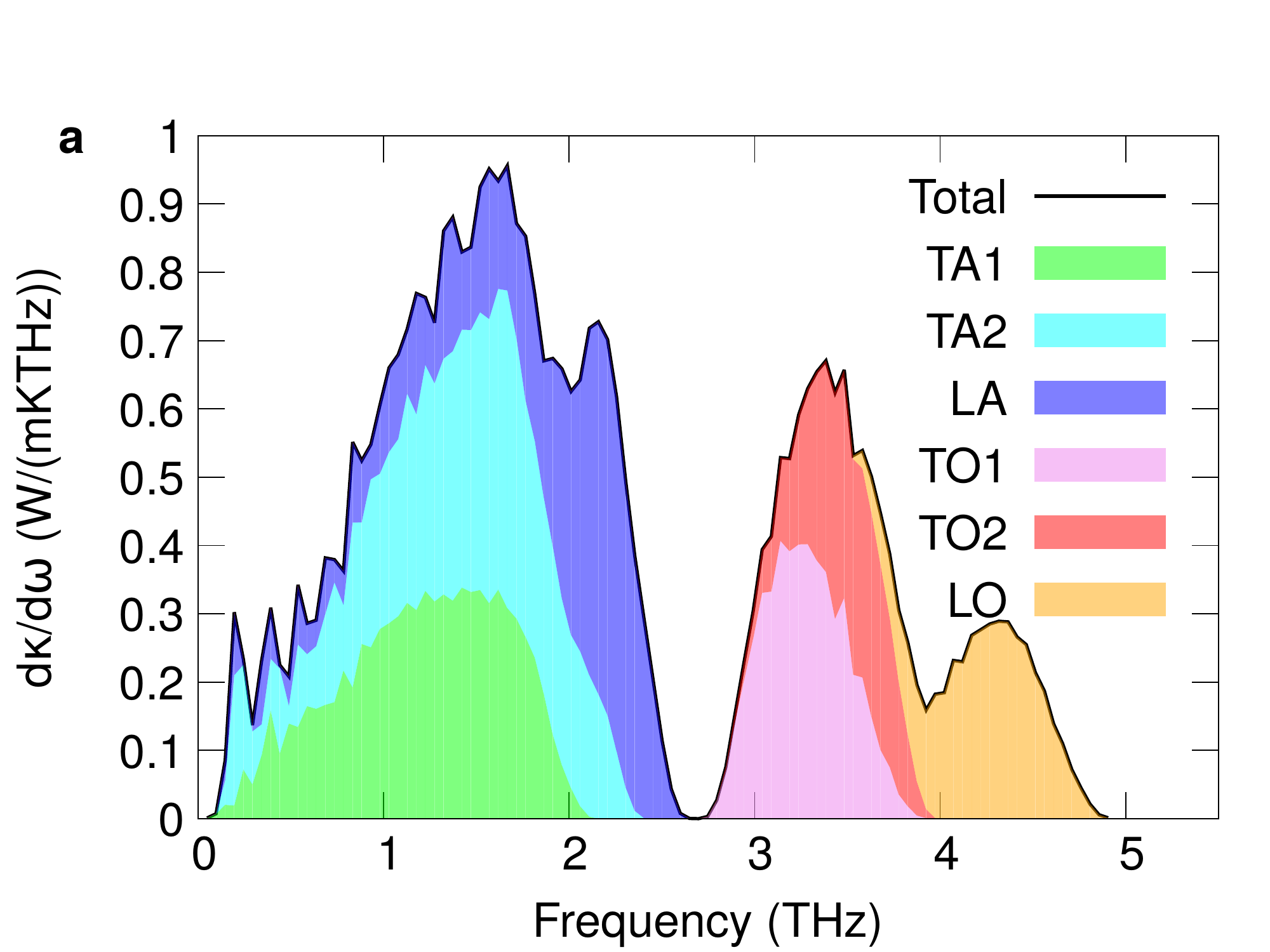}
\includegraphics[width=0.45\textwidth]{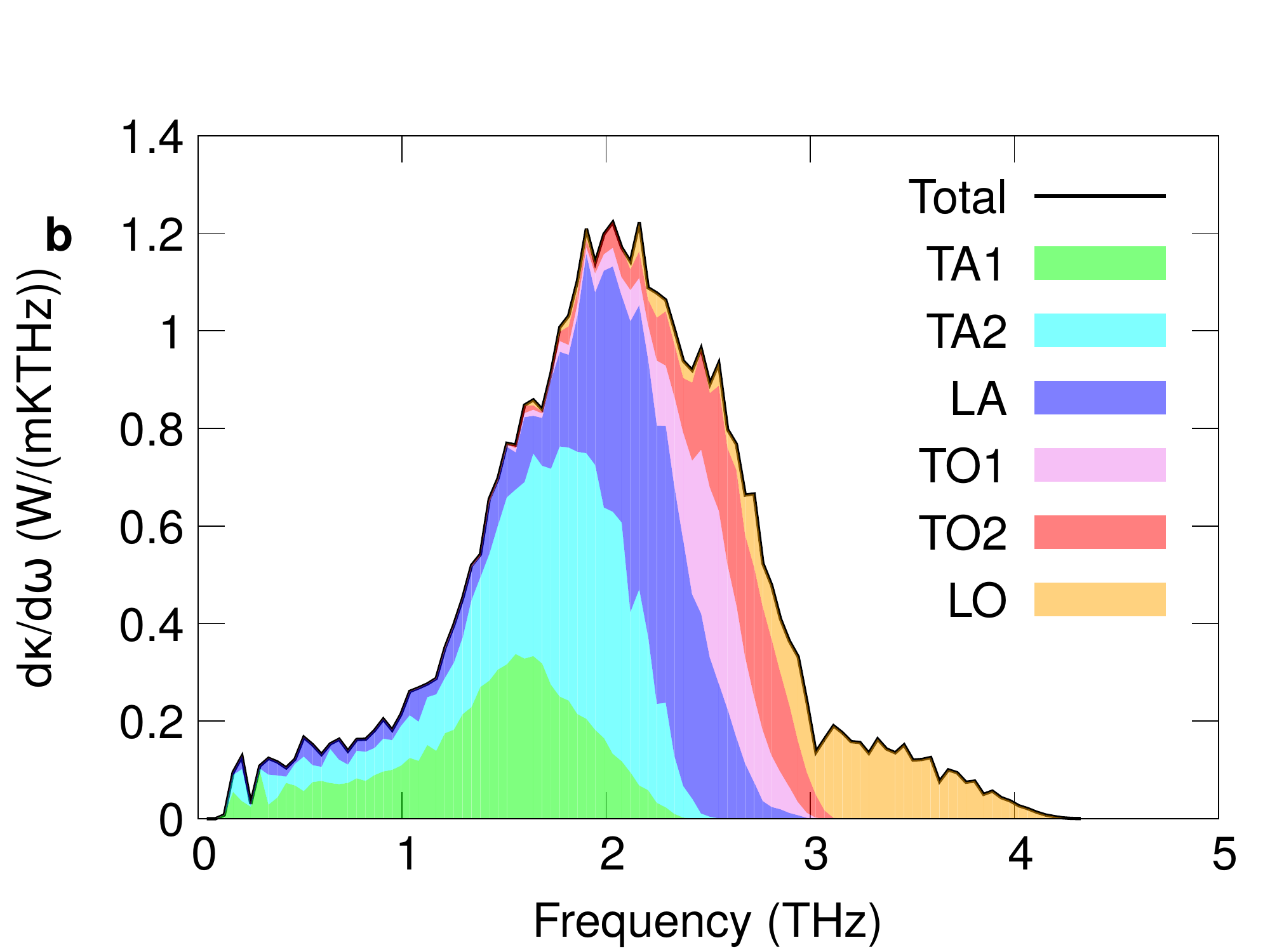}
\caption{Spectral lattice thermal conductivity of GeTe at \textbf{a} 300 K and \textbf{b} 631 K. Shaded regions show the contribution of a particular phonon branch.}
\label{supfig3}
\end{center}
\end{figure} 

\section{Supplementary note 5: Phonon linewidths of the Brillouin zone center optical modes}

We show comparison between the calculated and measured phonon linewidths \cite{SoftmodeGeTe} of the Brillouin zone center optical modes in Supplementary Figure~\ref{lw}. We extracted the theoretical linewidths by fitting lineshapes of phonon modes to the Lorentzian function. This method yielded similar results as taking the imaginary part of the self-energy at the anharmonic frequency and multiplying by 2. The use of this method (although probably identical to one used in the experiment, but further from the phase transition) is questionable at the phase transition and might be the source of numerical noise observed in the calculated values of linewidths, Supp. Fig.~\ref{lw}. 

\begin{figure}
\begin{center}
\includegraphics[width = 0.45\textwidth]{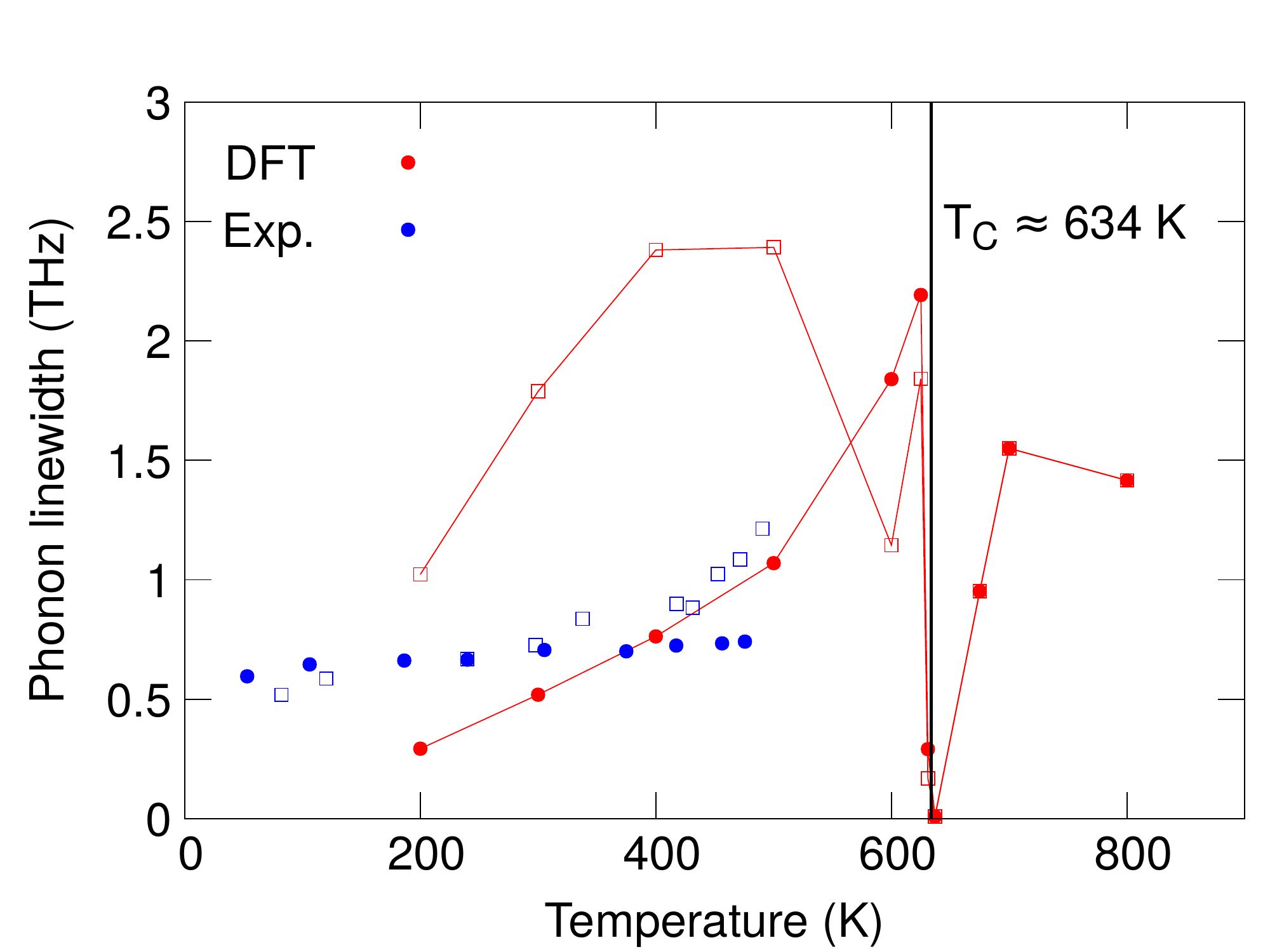}
\caption{Comparison between the measured and calculated phonon linewidths extracted from the phonon spectral functions. Red points correspond to our calculation, while blue points represent the measured values~\cite{SoftmodeGeTe}. Full circles denote the linewidths of the $E$ mode, while squares are the linewidths of A$_1$ mode.}
\label{lw}
\end{center}
\end{figure}

The theoretical results for phonon linewidths are remarkably noisy, with the soft mode (the A$_{1}$ mode as we approach $\Gamma$ from the $X-\Gamma$ direction) not showing a monothonic dependence on temperature. The E mode linewidth increases with temperature, with a peak at 625 K, and then plummets to close to the zero value at the phase transition (this is true for the A$_{1}$ mode as well). This is because the phonon spectral functions for these modes have peaks at 0 THz, which makes the extraction or the definition of linewidth troubling. The experiment shows the linewidths up to 500 K. Our theoretical results mostly overestimate the experimental phonon linewidths at higher temperatures.   

Comparing phonon lineshapes is more difficult due to the sheer amount of work needed to extract the data from the figures in Ref.~\cite{SoftmodeGeTe}. However, we can discuss the overall behavior of lineshapes. The dominant peaks in the experiment show only one peak per mode. This is true in our study for low temperatures (up to the maximum temperature in the experiment). The experiment has a large noise in the low frequency region, with unexplained peak structures. We do not observe those and they might not come from scattering of light from vibrational modes. Our results show a non-Lorentzian behavior even at low temperatures, but the resolution of the experiment is too low to show these details, if they exist.

\section{Supplementary Note 6: Phonon lifetimes close to the phase transition}

\begin{figure}
\begin{center}
\includegraphics[width=0.45\textwidth]{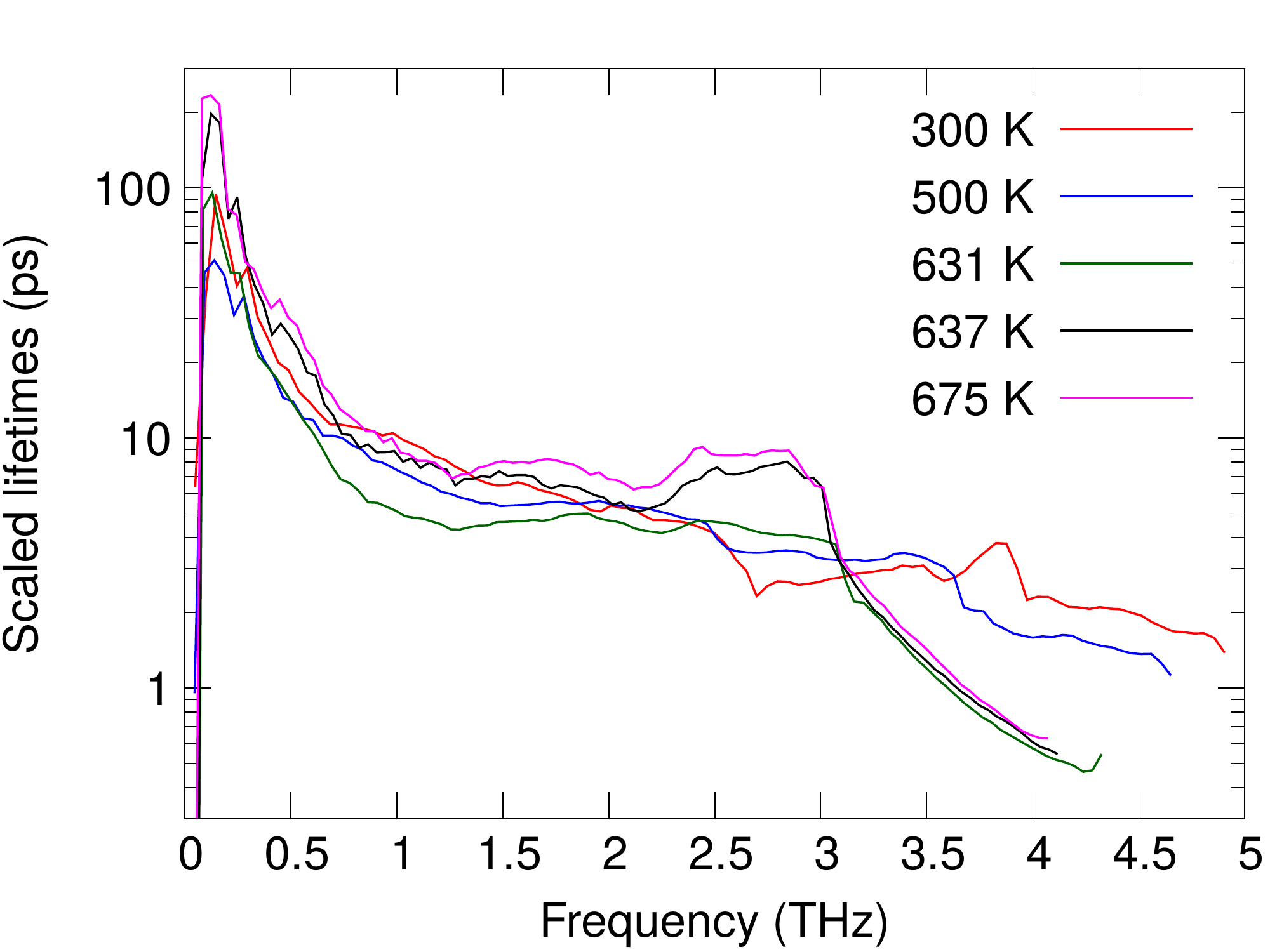}
\caption{Average phonon lifetimes of GeTe scaled by T/300 K, where T is the temperature. Averaging is carried out by convolving the calculated phonon lifetimes with a Gaussian.}
\label{supfig4}
\end{center}
\end{figure}

Supplementary figure \ref{supfig4} shows the average phonon lifetimes of GeTe at different temperatures scaled by $T/300$ K, where $T$ is the temperature. This enables us to see whether the decrease in the phonon lifetimes near the phase transition is due to phonon populations or increased anharmonicity of the material. Phonon lifetimes scale inversely with temperature for the temperatures far from the phase transition temperature ($T_{C} \approx 634$~K), which indicates that anharmonicity is not increased for those temperatures. However, close to the phase transition (631 K) in the rhombohedral phase, we can see a dip in the phonon lifetimes for most of the frequency range, revealing increased anharmonicity near the ferroelectric phase transition in the rhombohedral phase.

In the cubic phase, however, we see an increase in the scaled phonon lifetimes compared to the rhombohedral phase. Further from the phase transition (675 K), the scaled lifetimes are larger than the scaled phonon lifetimes at 300 K, which we attribute to lower intrinsic anharmonicity of the cubic phase.

To investigate in detail the reason for this behaviour of phonon lifetimes, we checked the change in anharmonic force constants with temperature (see Supplementary Fig.~\ref{supfig5}). To compare anharmonic force constants, we define a norm of the anharmonic force constant matrix as: $N^{AFC}(i,j,k) = \sum _{\alpha ,\beta ,\gamma} |\Phi ^{\alpha\beta\gamma}_{ijk}|$, where $i,j,k$ denote atoms in triplet and $\alpha , \beta ,\gamma$ are the Cartesian directions. Surprisingly, most of the anharmonic force constants decrease with temperature, in both rhombohedral and cubic phases. There is a sudden drop in the norm of the anharmonic force constants at the phase transition. The anharmonic force constants are drastically smaller in the cubic phase, explaining higher scaled phonon lifetimes in this phase.  

\begin{figure}
\begin{center}
\includegraphics[width=0.45\textwidth]{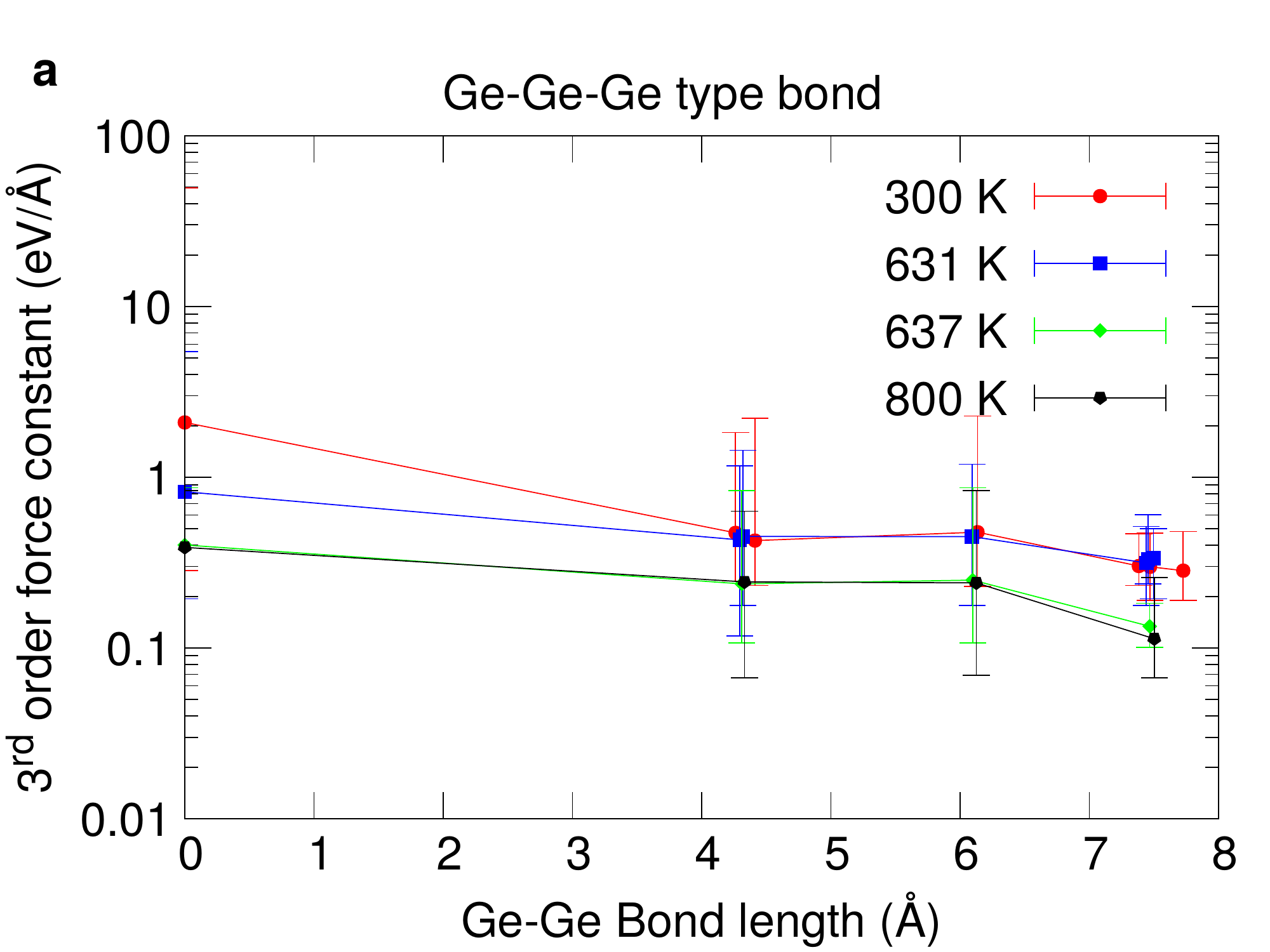}
\includegraphics[width=0.45\textwidth]{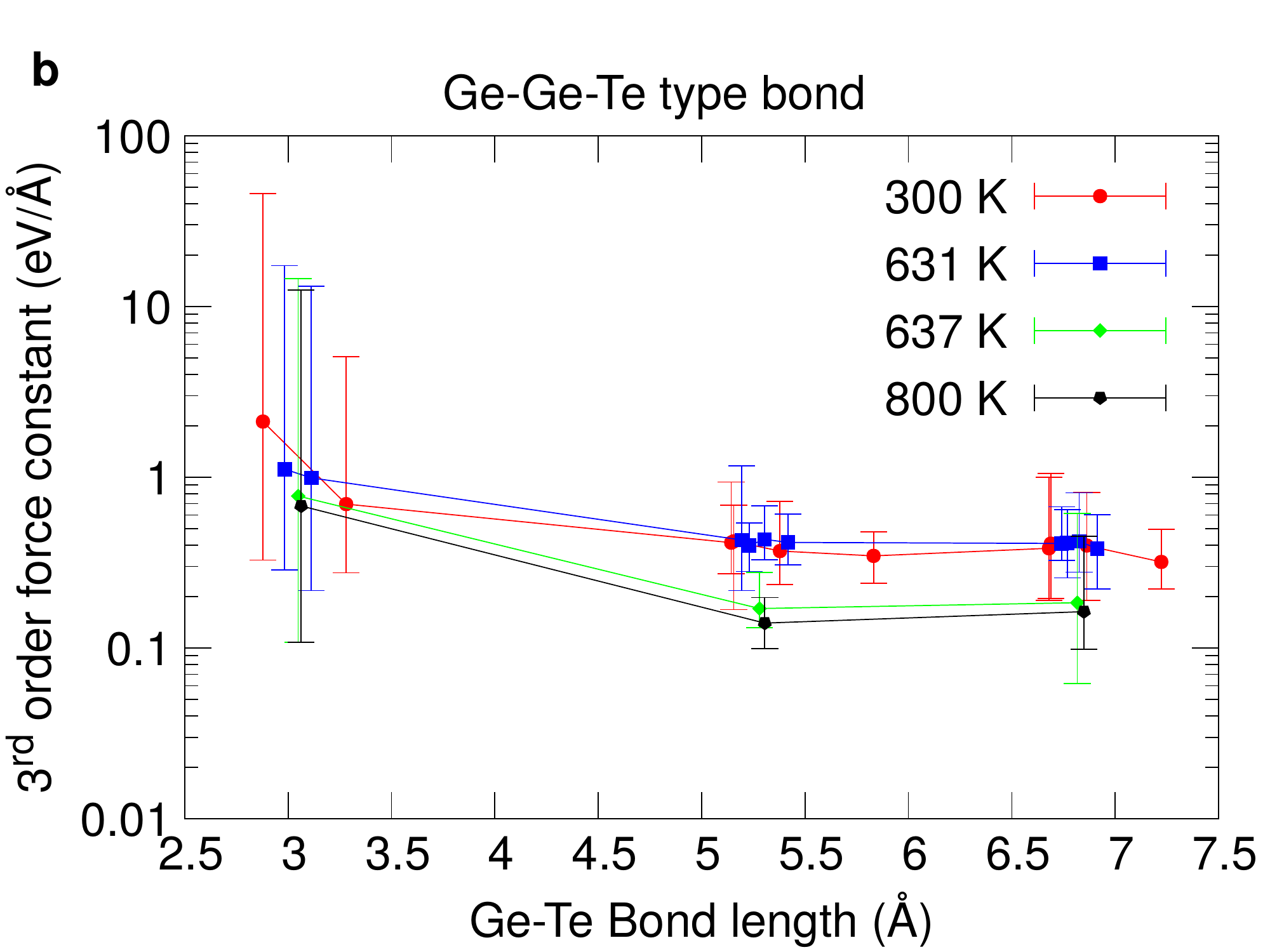}
\caption{Change of the anharmonic force constants of \textbf{a} Ge-Ge-Ge and \textbf{b} Ge-Ge-Te triplets with temperature. The bars in the figure show the maximum and minimum norm of the force constants for the specific temperature and bond length.}
\label{supfig5}
\end{center}
\end{figure} 

However, the temperature dependence of anharmonic force constants does not explain the increased anharmonicity of the transverse acoustic modes at the phase transition in the rhombohedral phase (see Supplementary Figure~\ref{supfig4}). It is the scattering phase space (SPS) that increases at the phase transition appreciably, which leads to higher anharmonicity in the rhombohedral phase. We show this in Supplementary Figure~\ref{supfig6} by calculating the scattering phase space for phonons at different temperatures. We do this by setting the matrix element $\Phi _{\lambda \lambda ' \lambda ''}$ in Eq. (11) of the main part to 1 and calculating the imaginary part of self-energy at the TDEP frequency for each phonon mode. The results are then scaled by temperature to minimize the effect of phonon population on the calculated value of the scattering phase space. We can notice that for the phonons in the frequency region around 1 THz the SPS increases dramatically near the phase transition, which explains the lower contribution of transverse optical modes to total $\kappa$ at the phase transition compared to 300 K (see Supplementary Figure~\ref{supfig3}) and prominent dip in the scaled phonon lifetimes for this frequency region (see Supplementary Figure~\ref{supfig4}). Phonons in the frequency region around 2 THz have a smaller SPS, which again is consistent with the results presented in Supplementary Figure~\ref{supfig4} (the scaled phonon lifetimes at the phase transition and 300 K are comparable in this frequency region). However, this can not be the reason for the increased lattice thermal conductivity at the phase transition, because this effect is noticeable only due to the temperature scaling of phonon lifetimes and SPS and does not exist if one takes phonon populations into account (see Fig. 5 \textbf{b} of the main part). Finally, the available scattering phase space of longitudinal optical (the highest frequency) phonons increases dramatically with temperature, which explains their unusual frequency dependence. 

In the cubic phase the SPS of the majority of phonons (except LO phonons) increases linearly with temperature, which balances out the decrease in the anharmonic force constants and leads to almost constant phonon lifetimes with temperature. The decrease in $\kappa$ in the cubic phase between 700 K and 800 K comes primarily from a decrease in phonon group velocities. 

\begin{figure}
\begin{center}
\includegraphics[width=0.45\textwidth]{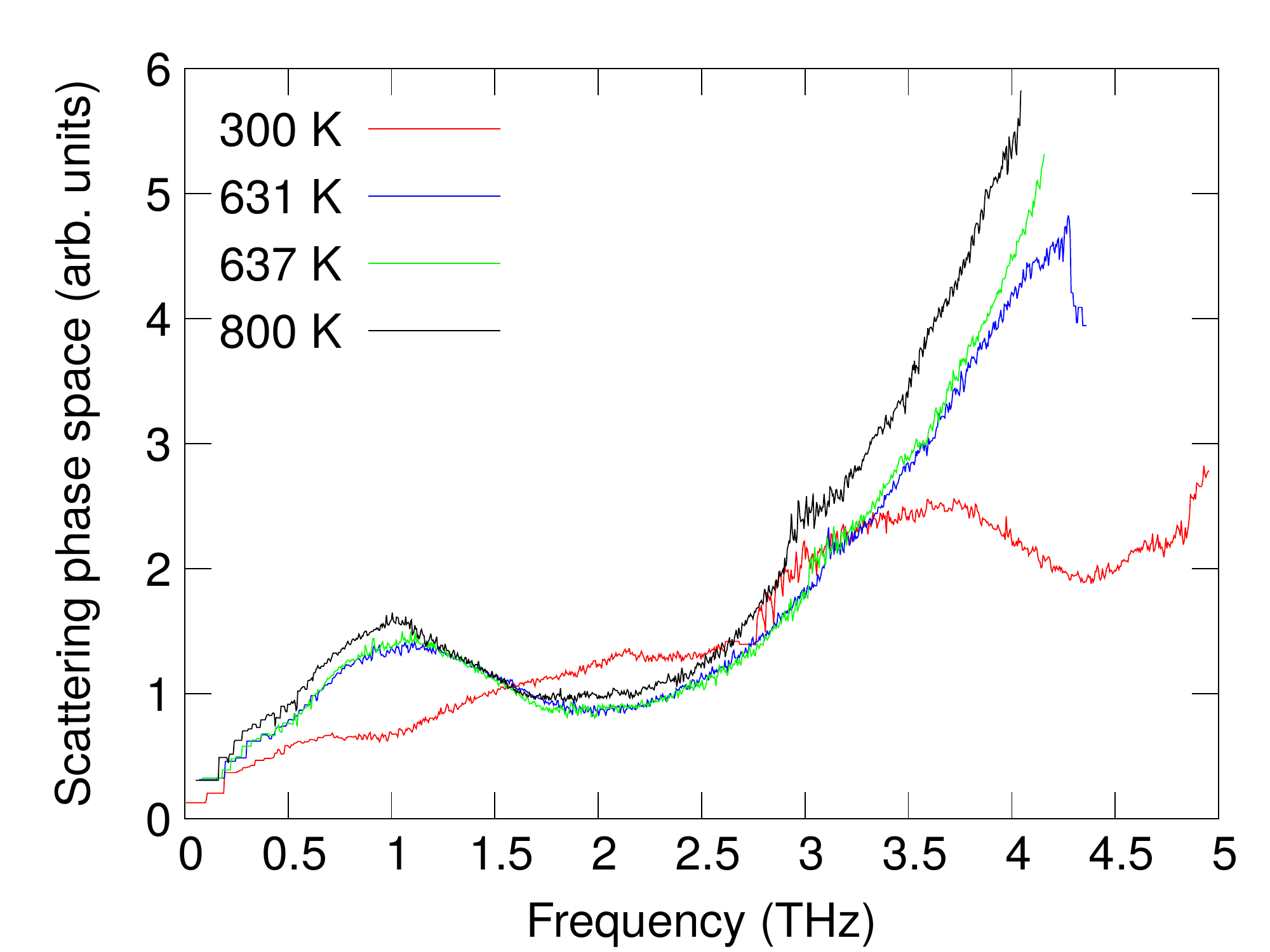}
\caption{Phonon scattering phase space scaled by T/300 as a function of TDEP frequency at different temperatures.}
\label{supfig6}
\end{center}
\end{figure} 

\section*{Supplementary note 7: Gaussian Approximation Potential}

To accurately sample atomic displacements at a certain temperature for the temperature dependent effective potential (TDEP) method, one must perform molecular dynamics (MD) simulations. Most commonly, one would run ab initio MD to achieve this. However, ab initio MD is extremely computationally expensive and the accuracy of such calculations is typically reduced in order to speed up the computation of forces and energies. Another important drawback of ab initio  MD is that it does not scale linearly with the number of atoms, so carrying out these calculations on large supercells needed to reach convergence in GeTe would be very expensive.

To overcome these difficulties, we developed an accurate interatomic potential based on the Gaussian Approximation Potential (GAP) framework using ab initio (density functional theory, DFT, with the GGA-PBE exchange-correlation functional) calculated forces for different atomic configurations in different GeTe supercells. We fitted the GAP potential using an iterative procedure as follows. We created the first training set of atomic forces and energies using randomly displaced atoms at different temperatures for different sizes of supercells (128 - 512 atoms). The temperature for these structures was defined through the average square of atomic displacements from equilibrium positions. We trained an initial version of GAP and ran a MD simulation using it as a force field. The MD simulation fails at some temperature (some atoms would leave a simulation box) and we sample a number of atomic configurations along that MD trajectory. We recalculate forces and energies for these structures in DFT and include them into the training set of the potential. We refit GAP with an updated training set. We repeat these steps (running MD, sampling configurations along the MD trajectory, adding them to the training set and training a new version of GAP) until we reach a sufficiently converged version of the interatomic potential. By this we mean that the MD simulations are stable for a range of temperatures and structures, and the potential does not show a significant improvement upon adding new structures to the training set. The hyperparameters of the GAP model (we used only Smooth Overlap of Atomic Positions-SOAP kernel) are given in Table~\ref{tb3} ~\cite{GAP1, GAP2}.

\begin{table*}[h]
\begin{center}
\begin{tabularx}{0.95\textwidth}{| Y | Y | }
\hline \hline
 Cutoff radius            & 6.5 \AA   \\
 \hline
 Smooth cutoff transition & 0.8 \AA  \\
 \hline
 Energy regularization    & 0.001 eV per atom \\
 \hline
 Force regularization     & 0.02 eV \AA $^{-1}$  per atom \\
 \hline
 Kernel exponent          & 4 \\
 \hline 
 Sparse jitter            & 10$^{-8}$ \\
 \hline 
 (n$_{\text{max}}$, l$_{\text{max}}$) & (6,8) \\
 \hline \hline
\end{tabularx}
\end{center}
\caption{Gaussian Approximation Potential hyperparameters~\cite{GAP1, GAP2} used to fit the interatomic potential for GeTe.} 
\label{tb3}
\end{table*} 

To justify the use of this interatomic potential, we have prepared some comparisons between the GAP potential and DFT. In Table~\ref{tb1} we show the comparison of the GAP 0 K structure with DFT using different exchange-correlation functionals and experiment \cite{main}. We fitted the GAP potential to DFT with the GGA-PBE functional calculations. We see that the agreement of GAP with DFT is satisfactory.

\begin{table*}[h]
\begin{center}
\begin{tabularx}{0.95\textwidth}{ c | Y | Y | Y | Y }
\hline \hline
  & $a$ (\r{A}) &$\theta$ (deg)&$\mu $&$V_{0}$ (\r{A}$ ^{3}$) \\
 \hline
  \textsc{LDA}      & 4.207 & 58.788  & 0.024 & 51.193 \\ 
 \hline 
  \textsc{GGA-PBE}  & 4.381 & 57.776  & 0.030 & 56.420   \\
 \hline
  \textsc{GAP}      & 4.422 & 56.861  & 0.032 & 56.700 \\
 \hline
 Experiment (295 K) & 4.299 & 57.931  & 0.025 & 53.513  \\
 \hline
 GAP (300 K)        & 4.453 & 56.850  & 0.032 & 57.903 \\
 \hline \hline
\end{tabularx}
\end{center}
\caption{Lattice parameters of GeTe at 0 K, calculated using the interatomic potential, DFT with the \textsc{LDA} and \textsc{GGA-PBE} functionals, and compared with the experimental results at 295 K and the GAP results at 300 K. $a$ stands for lattice constant, $\theta$ for angle, and $\mu$ for internal atomic coordinate.} 
\label{tb1}
\end{table*} 

To further justify the use of GAP, we compare the elastic constants ($\hat{\text{C}}$) and bulk modulus (B) calculated using GAP and density functional perturbation theory (DFPT) with the GGA-PBE functional. Here we will report the ion relaxed values of these quantities. Bulk modulus was calculated using the Voigt average. As we can see from Table \ref{tb5_2}, all of the calculated values agree in the range of 20\% error. The accuracy with which we determined elastic constants assures that our results regarding lattice thermal expansion will be sound. 

\begin{table*}[h]
\begin{center}
\begin{tabularx}{0.8\textwidth}{ c | X | X | X | X | X | X | X }
\hline \hline
 (GPa) & C$_{11}$ &C$_{12}$ &C$_{13}$ &C$_{33}$ &C$_{14}$ &C$_{44}$ & B  \\
 \hline
  \textsc{GAP}      & 79 & 21  & 20 & 30 & 14 & 21 & 34 \\ 
 \hline 
  \textsc{GGA-PBE}  & 87 & 18  & 18 & 36 & 15 & 25 & 35   \\
 \hline \hline
\end{tabularx}
\end{center}
\caption{Elastic constants of germanium telluride calculated using the \textsc{GAP} interatomic potential and DFPT with the \textsc{GGA-PBE} exchange-correlation functional.} 
\label{tb5_2}
\end{table*} 

Further, we compared the phonon band structures calculated using forces from DFT and GAP (see Supplementary Fig.~\ref{supfig7} \textbf{a}). The comparison between these two methods is very good. 

To further check the appropriateness of GAP for MD simulations, we compared the errors from GAP and DFT. We calculate forces on a set of structures and atomic configurations using DFT with fully converged parameters. Then we calculate the same forces using GAP and find the differences with respect to DFT. We then bin the errors and present the results in Supplementary Fig.~\ref{supfig7} \textbf{b}. We highlight that the structures taken in this study are not in the training set used to obtain the GAP potential which makes the low value of errors in forces even more remarkable.

\begin{figure*}
\begin{center}
\includegraphics[width=0.45\textwidth]{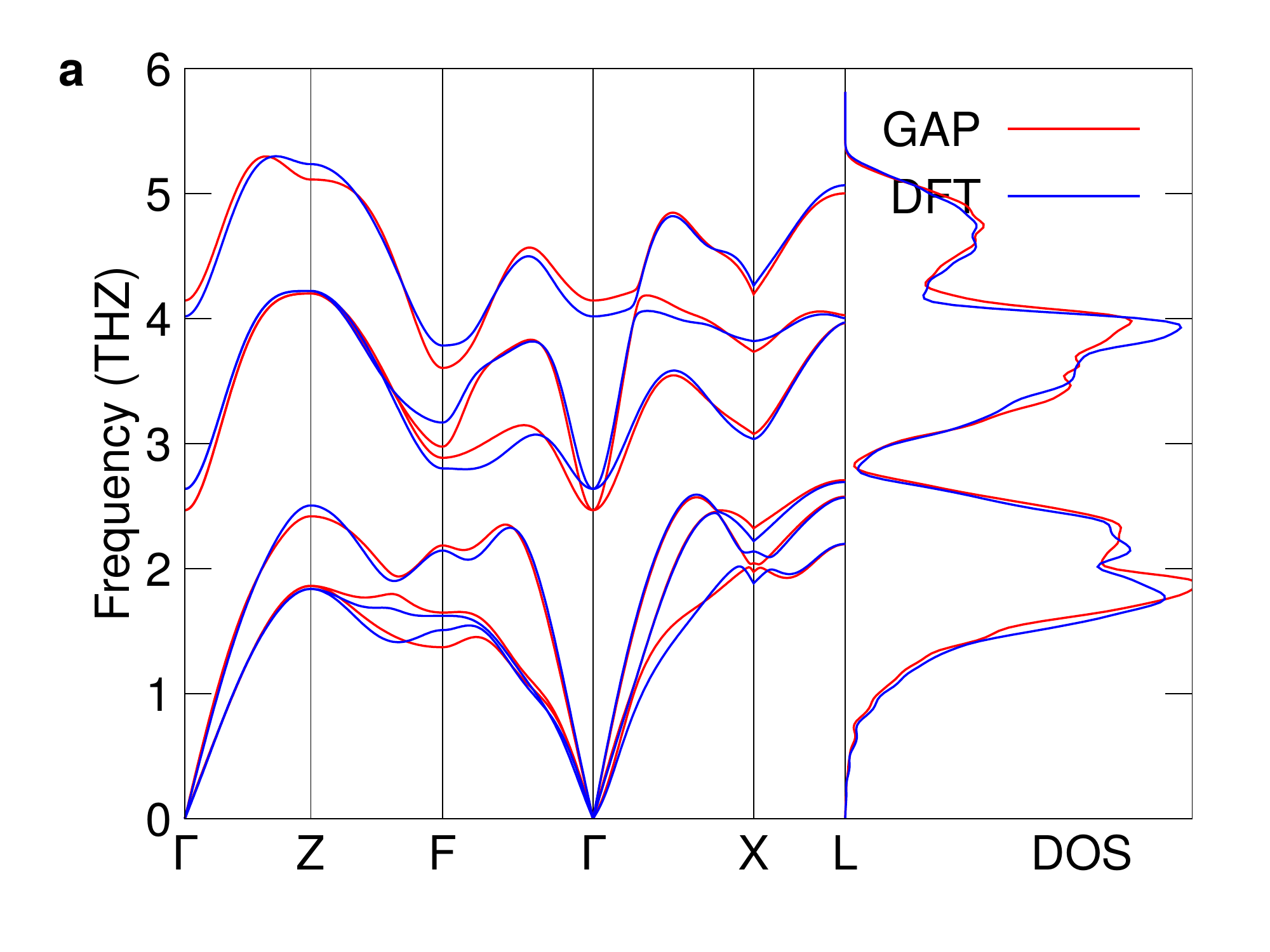}
\includegraphics[width=0.45\textwidth]{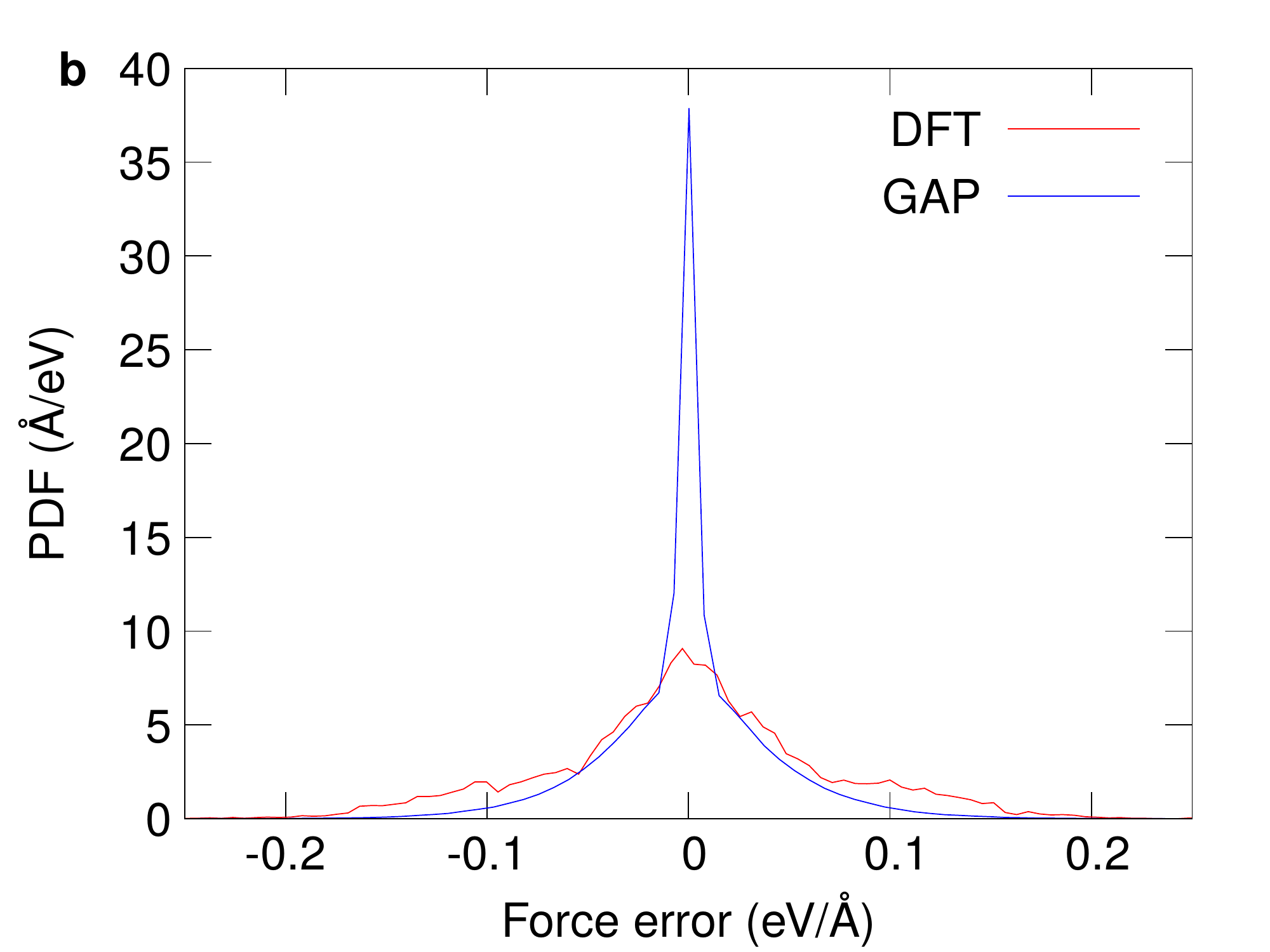}
\includegraphics[width=0.45\textwidth]{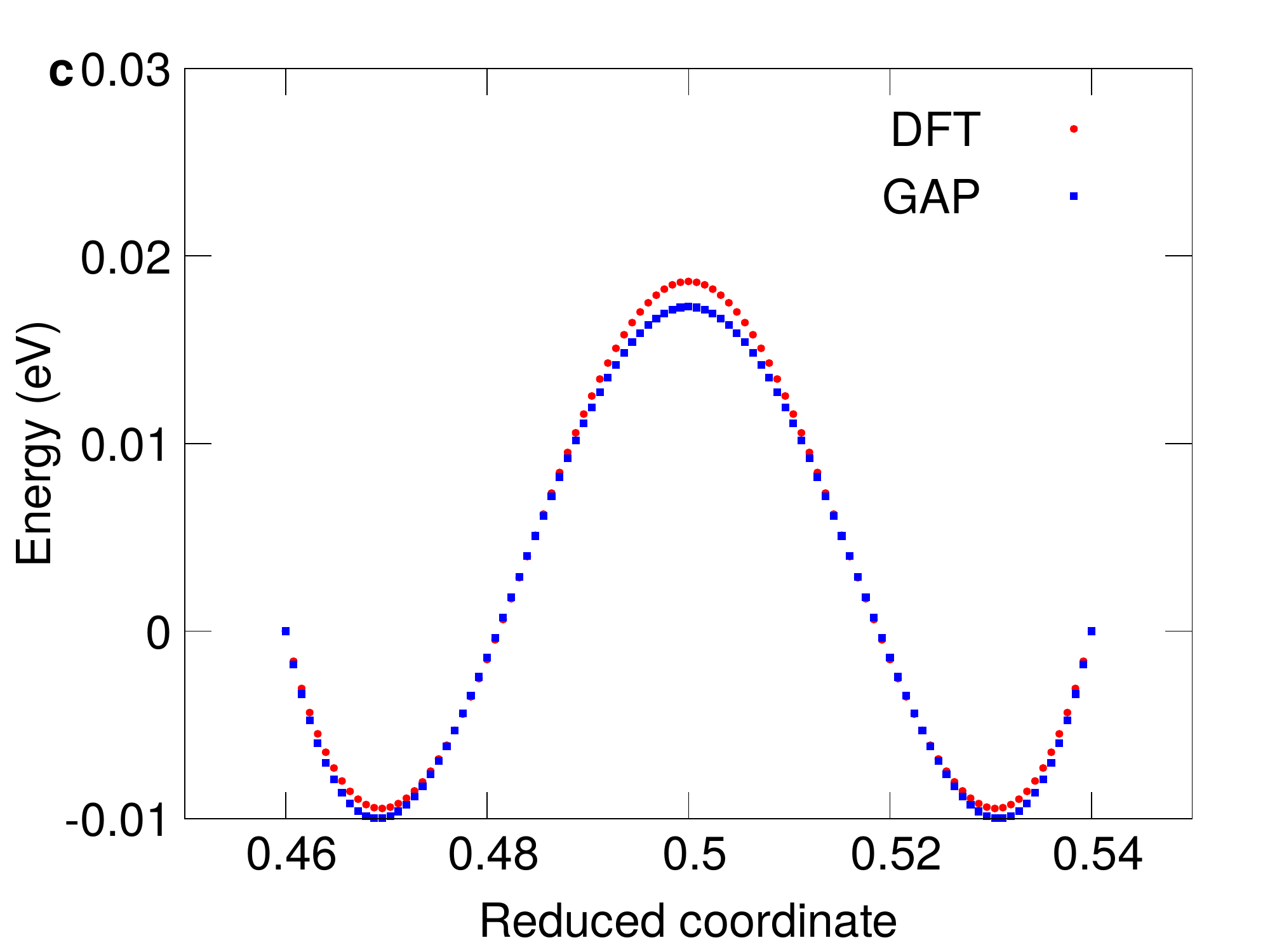}
\caption{\textbf{a} Phonon band structure of GeTe at 0 K calculated using forces from density functional theory (DFT) and Gaussian Approximation Potential (GAP). \textbf{b}  The force error probability distribution function for GAP and DFT with reduced accuracy (see text for more details). \textbf{c} The energy profile of the interatomic displacement parameter calculated using GAP and DFT.}
\label{supfig7}
\end{center}
\end{figure*} 

As we previously said, to actually run ab initio MD simulations, one would have to reduce the accuracy of DFT calculations. To mimic this, we calculated forces on 216 atom supercells of GeTe using a 2x2x2 $\vec{k}$-point grid and the energy cutoff of 16 Ha for plane waves. We consider this calculation as high accuracy. To perform low accuracy calculations, we chose an 1x1x1 $\vec{k}$-point grid and the energy cutoff of 10 Ha for the plane wave expansion of electronic wave functions. We then take the differences in the forces in these two calculations and bin them in a similar manner to the previous case with the GAP potential (see Supplementary Fig.~\ref{supfig7} \textbf{b}). We can see that the errors from the GAP potential are comparable to the errors obtained from the reduced accuracy first principles calculations, which gives us confidence in the sampling method we used. 

Finally, Supplementary figure~\ref{supfig7} \textbf{c} shows the comparison of interatomic displacement parameter energy profile between GAP and DFT. We fixed the values of lattice constant and rhombohedral angle at the same value for the GAP and DFT (0 K DFT GGA-PBE structure).We can see excellent agreement between two approaches. 

\section{Supplementary Note 8: Structural parameters of GeTe}

\begin{figure*}
\begin{center}
\includegraphics[width = 0.45\textwidth]{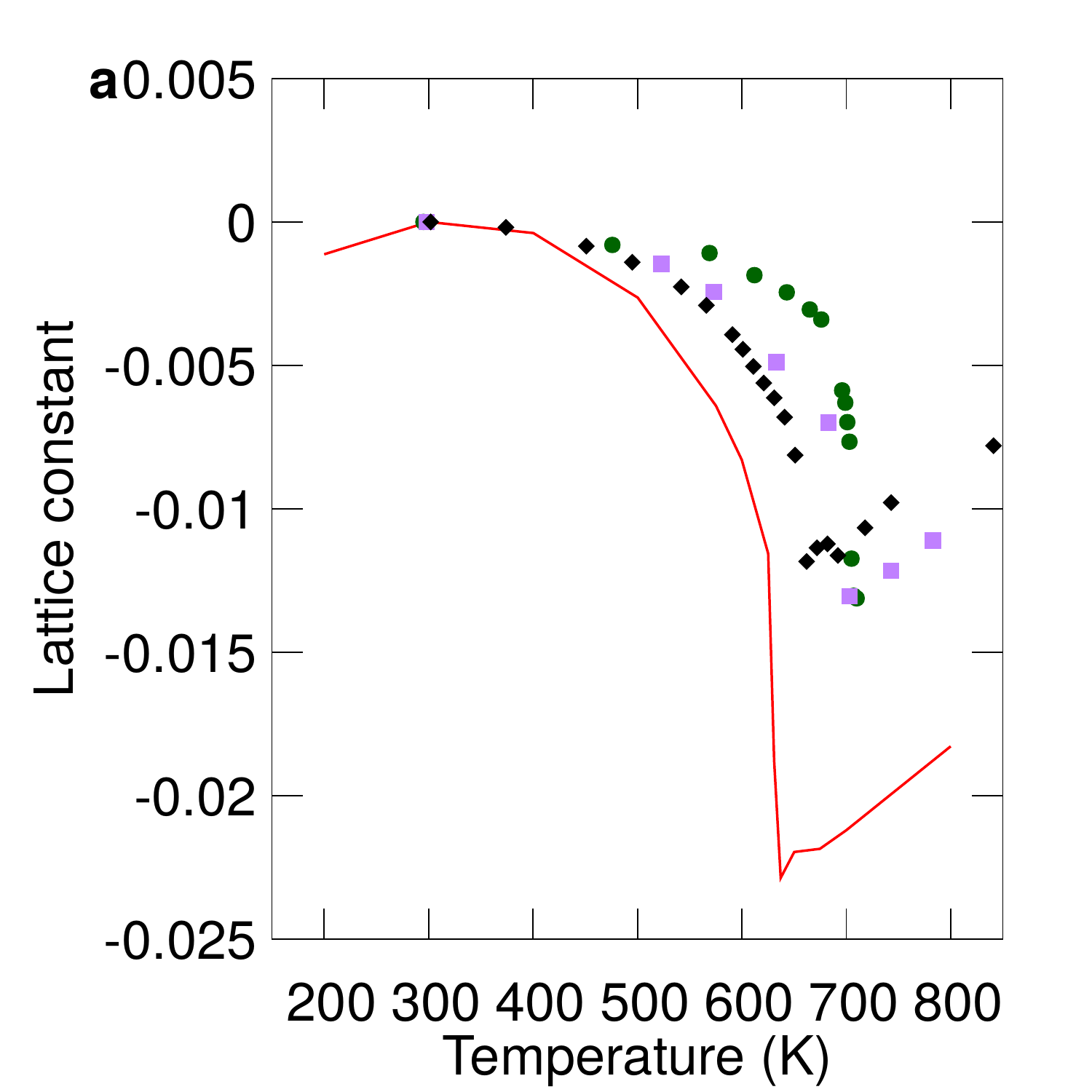}
\includegraphics[width = 0.45\textwidth]{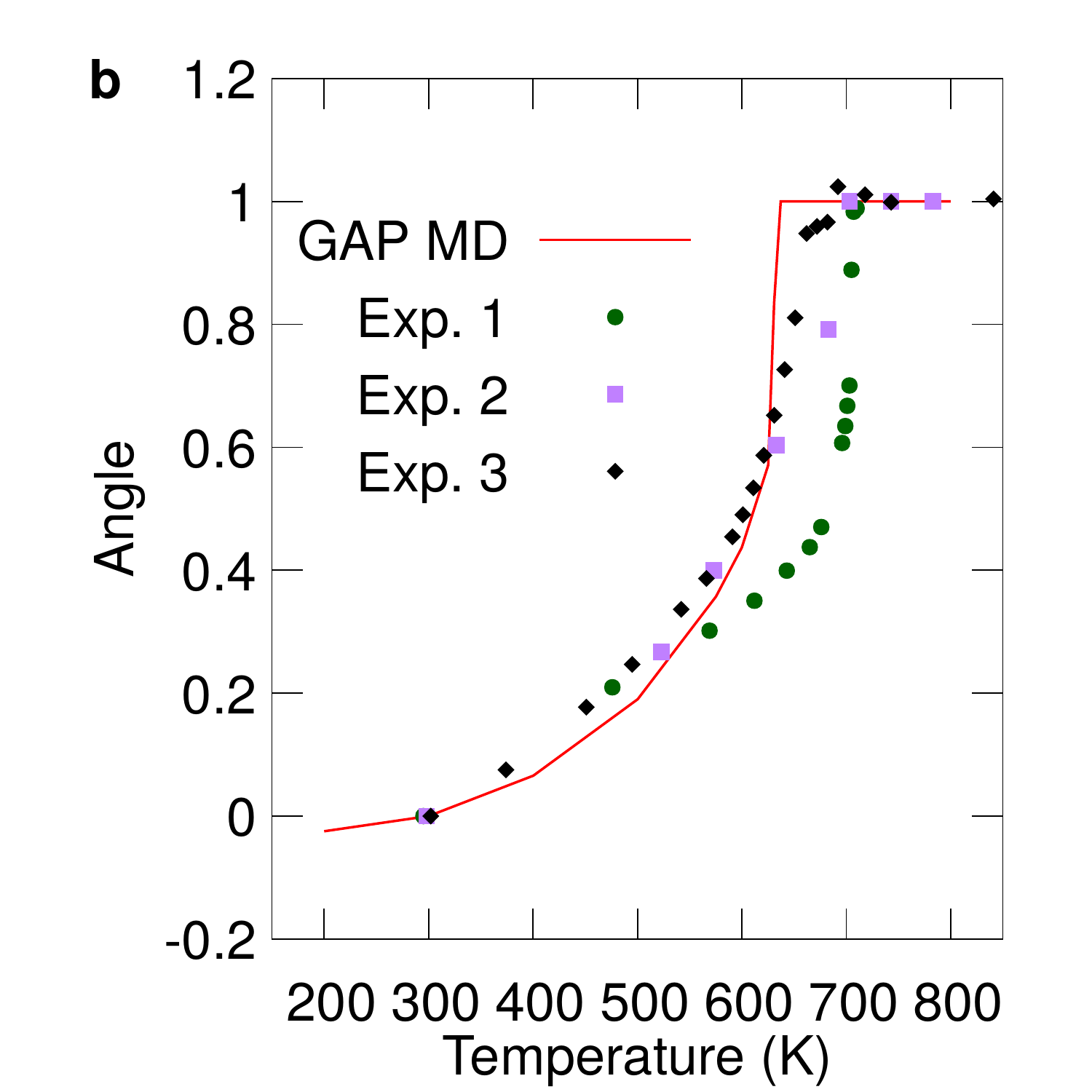}
\includegraphics[width = 0.45\textwidth]{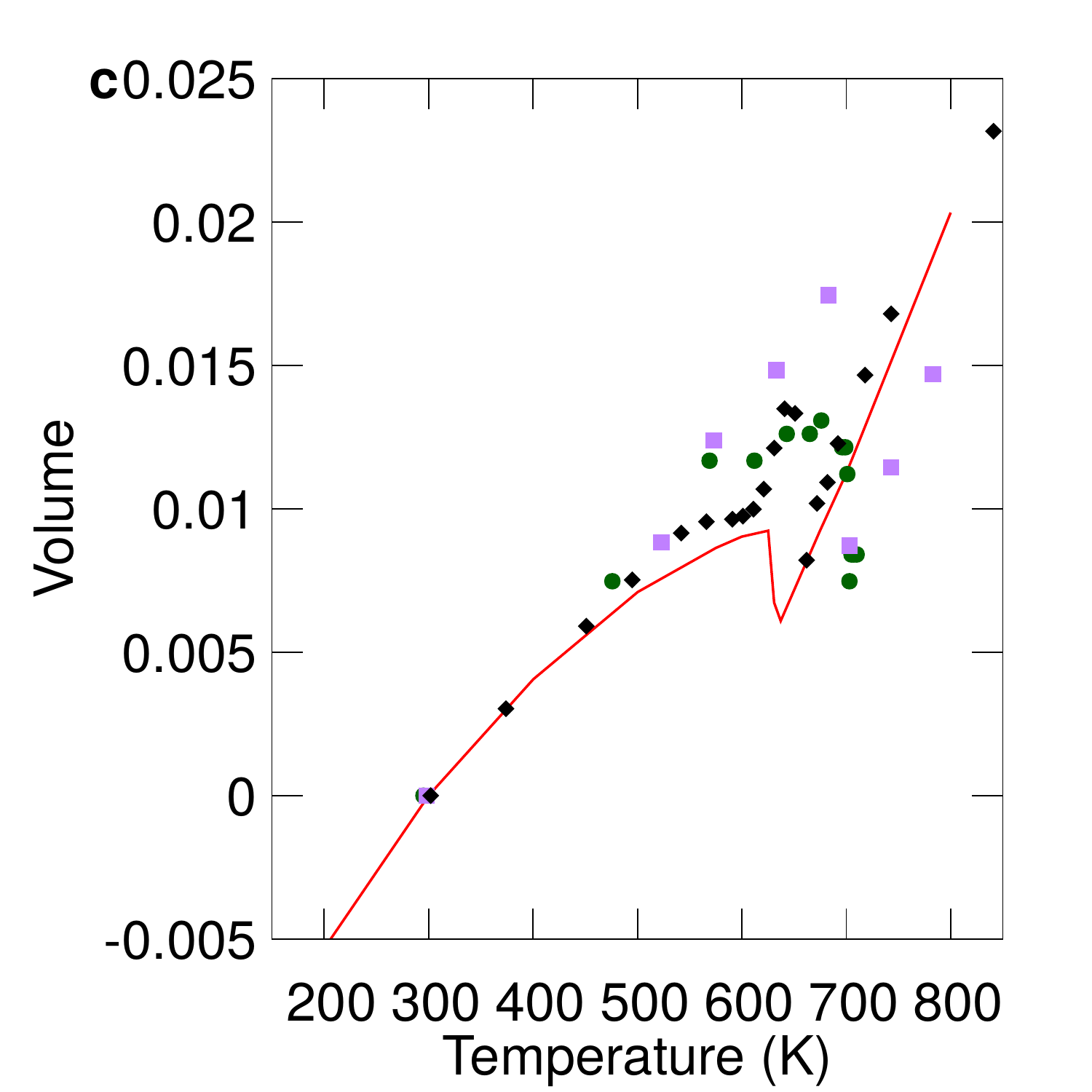}
\includegraphics[width = 0.45\textwidth]{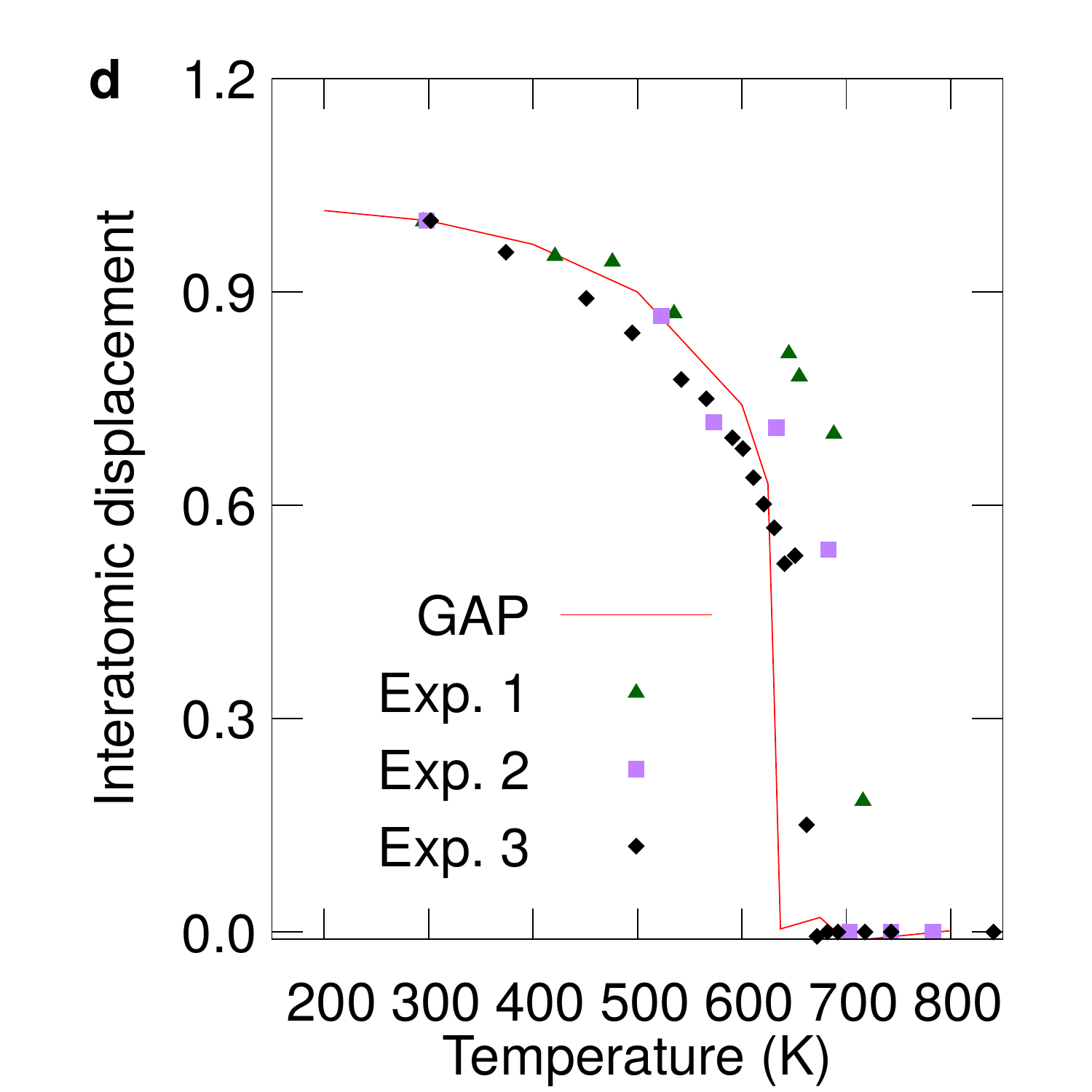}
\caption{Thermal variation of \textbf{a} lattice constant, \textbf{b} rhombohedral angle, \textbf{c} volume, and \textbf{d} interatomic displacement parameter in germanium telluride using molecular dynamics simulation compared to experiments. The standard errors associated with the set of values for structural parameters obtained along MD trajectory are not visible on the graph (smaller than experimental points). Experiments correspond to Refs.\cite{main, GeTeBo, mdpigete}}
\label{supfig9}
\end{center}
\end{figure*}

The thermal expansion coefficients for the structural parameters of GeTe obtained using the described MD simulations with the GAP potential are presented in Supplementary Fig.~\ref{supfig9} in comparison with experiments \cite{main, GeTeBo, mdpigete}. Here we show the following quantities ($\alpha$):
\begin{align*}
\alpha _{u} &= \frac{u_T - u_{300}}{u_{300}} \quad \text{for } u = a, V , \\
\alpha _{\theta} &= \frac{\theta _T - \theta _{300}}{60 - \theta _{300}}, \\
\alpha _{\mu} &= \frac{\mu _T}{\mu _{300}}.
\end{align*}
However, the density functional theory (DFT) and the GAP global minimum of the Born-Openheimer energy surface differ slightly (see Table ~\ref{tb1}). This is why we extract the relative change of structural parameters from the MD study of thermal expansion of GeTe using GAP and use it to infer the values of structural parameters at finite temperature for the DFT study. This is equivalent to saying that we obtained thermal expansion coefficients from MD simulations using GAP and then used them to obtain the finite temperature structure of GeTe in DFT.

The results for the interatomic displacement parameter $\mu$ in Supplementary Fig.~\ref{supfig9} (d) clearly show that $\mu$ has a non-zero value at 631 K and a negligible value at 637 K. Therefore, we can consider that GeTe at 637 K is already in the cubic, paraelectric, phase. That would mean that the phase transition happens between 631 K and 637 K. We took the average of these two temperatures to estimate the critical temperature in this system ($T_{\text{C}} = 634 K$).

The results for volume expansion in Supplementary Fig.~\ref{supfig9} (c) show that GeTe has negative thermal expansion at the phase transition. This is in agreement with our previous study on lattice thermal expansion of GeTe \cite{Our}, where we used DFT and developed a model based on the Gr{\"u}neisen theory of thermal expansion to calculate the structural parameters of GeTe. We also obtained that negative thermal expansion occurs near the ferroelectric phase transition in GeTe. In addition to the results given in Supplementary Fig.~\ref{supfig9}, we also give the structural parameters values at 300 K obtained using the GAP potential in Table~\ref{tb1}.

\end{document}